\DocumentMetadata{uncompress} 
\documentclass[
  10pt,
  longbibliography,
  twocolumn,
  twoside,
]{article}

\usepackage[numbers,round,comma,sort&compress]{natbib}

\usepackage[
  left=0.65in,
  right=0.65in,
  top=0.65in,
  bottom=0.65in,
]{geometry}

\usepackage{sectsty}
\sectionfont{\raggedright\large\bfseries}
\subsectionfont{\raggedright\large}

\usepackage[font=small,labelfont=bf]{caption}

\usepackage{fancyhdr}
\usepackage[realmainfile]{currfile}

\newcommand{\onlinesiteshort}{compstorylab.org/archetypometrics}
\newcommand{\onlinesite}{https://\onlinesiteshort}

\usepackage{amssymb}
\usepackage{amsmath}

\usepackage{url}
\PassOptionsToPackage{hyphens}{url}\usepackage{hyperref}
\makeatletter
\g@addto@macro{\UrlBreaks}{\UrlOrds}
\makeatother

\usepackage[table]{xcolor}   
\usepackage{colortbl}

\definecolor{goodblue}{RGB}{0, 91, 187}
\usepackage{hyperref}
\hypersetup{
  colorlinks=true,
  allcolors=goodblue,
  urlcolor=goodblue,
  citecolor=goodblue,
  pdfborder={0 0 0},
  breaklinks=true,
}

\usepackage{stmaryrd}

\newcommand{\externallinksymbol}{{\tiny$^{{}_{\nnearrow}}$\!\!}}
\newcommand{\internallinksymbol}{{$^{\Rsh}$\!\!}}
\newcommand{\paperlinksymbol}{\externallinksymbol}

\usepackage[normalem]{ulem}

\usepackage{textcomp}

\usepackage[export]{adjustbox}

\makeatletter

\def\CT@@do@color{%
  \global\let\CT@do@color\relax
  \@tempdima\wd\z@
  \advance\@tempdima\@tempdimb
  \advance\@tempdima\@tempdimc
  \advance\@tempdimb\tabcolsep
  \advance\@tempdimc\tabcolsep
  \advance\@tempdima2\tabcolsep
  \kern-\@tempdimb
  \leaders\vrule
  \hskip\@tempdima\@plus  1fill
  \kern-\@tempdimc
  \hskip-\wd\z@ \@plus -1fill }
\makeatother

\usepackage[table]{xcolor}

\definecolor{olivegreen}{rgb}{0.33333,.41961,0.18431}
\definecolor{forestgreen}{rgb}{0.13333,.5451,0.13333}

\definecolor{lightgrey}{rgb}{0.7,0.7,0.7}
\definecolor{verylightgrey}{rgb}{0.90,0.90,0.90}
\definecolor{veryverylightgrey}{rgb}{0.95,0.95,0.95}
\definecolor{grey}{rgb}{0.5,0.5,0.5}
\definecolor{darkgrey}{rgb}{0.3,0.3,0.3}
\definecolor{verydarkgrey}{rgb}{0.15,0.15,0.15}

\definecolor{headerblue}{HTML}{33367E}
\definecolor{unitednationsblue}{HTML}{4D88FF}

\definecolor{charcoal}{HTML}{36454F}
\definecolor{cinerous}{HTML}{98817B}
\definecolor{feldgrau}{HTML}{4D5D53}
\definecolor{glaucous}{HTML}{6082B6}
\definecolor{arsenic}{HTML}{3B444B}
\definecolor{xanadu}{HTML}{738678}

\definecolor{firebrick}{HTML}{B22222}
\definecolor{orangered}{HTML}{FF4500}
\definecolor{tomato}{HTML}{FF6347}

\definecolor{orange}{RGB}{255,116,0}

\definecolor{purpletaupe}{HTML}{3B444B}

\definecolor{rose}{HTML}{E3242B}

\colorlet{editnotecolor}{rose}

\definecolor{headerorange}{RGB}{255,116,0}
\definecolor{headergray}{RGB}{230,230,230}

\definecolor{headerpop}{RGB}{230,230,230}

\definecolor{magmalight}{RGB}{252,251,195}
\definecolor{magmalightalt}{RGB}{250,240,184}
\definecolor{magmamedium}{RGB}{245,200,146}
\definecolor{magmadark}{RGB}{224,106,98}

\definecolor{icelight}{RGB}{223,242,244}
\definecolor{icelightalt}{RGB}{189,222,226}
\definecolor{icemedium}{RGB}{132,184,204}
\definecolor{icedark}{RGB}{103,153,191}

\definecolor{traitrowcolor}{RGB}{223,242,244}
\definecolor{traitrowcoloralt}{RGB}{189,222,226}

\definecolor{characterrowcolor}{RGB}{252,251,195}
\definecolor{characterrowcoloralt}{RGB}{250,240,184}

\definecolor{archetyperowcolor}{RGB}{255,213,212} 
\definecolor{archetyperowcoloralt}{RGB}{255,182,179} 

\definecolor{datasetrowcolor}{RGB}{232,244,234}
\definecolor{datasetrowcoloralt}{RGB}{210,231,214}

\newcommand{\done}[1]{}

\usepackage{titlesec}
\titleformat*{\paragraph}{\bfseries}

\usepackage{changepage}

\usepackage{fancyvrb}

\usepackage{tcolorbox}
\tcbuselibrary{skins,breakable,listings,breakable}

\usepackage{tikz}
\usetikzlibrary{shapes}

\tikzstyle{mybox} = [draw=lightblue!70, fill=lightblue!7, very thick,
    rectangle, rounded corners, inner sep=10pt, inner ysep=20pt]

\tikzstyle{editortitle} =[draw=archetyperowcoloralt, fill=archetyperowcoloralt, text=black]

\newcommand\Loadedframemethod{default}
\usepackage[framemethod=\Loadedframemethod]{mdframed}

\mdfsetup{skipabove=\topskip,skipbelow=\topskip}

\tikzstyle{loglinetitle} =[draw=icedark, fill=icemedium!50, text=black]

\newenvironment{loglinebox}[1][]{

  \ifstrempty{#1}%
  {\mdfsetup{%
    frametitle={%
       \tikz[baseline=(current bounding box.east),outer sep=0pt]
        \node[loglinetitle, anchor=east,rectangle]
        {\strut~~#1~~\strut};}}
  }%
  {\mdfsetup{%
     frametitle={%
       \tikz[baseline=(current bounding box.east),outer sep=0pt]
        \node[loglinetitle,anchor=east,rectangle]
        {\strut~~#1~~\strut};}}%
   }%
   \mdfsetup{innertopmargin=5pt,linecolor=icedark,%
             linewidth=0.5pt,topline=true,
             frametitleaboveskip=\dimexpr-\ht\strutbox\relax,}
   \begin{mdframed}[backgroundcolor=icelight,nobreak=true]\relax%
     \raggedright
}{\end{mdframed}}

\tikzstyle{abstracttitle} =[draw=magmadark!75, fill=magmamedium!75, text=black]

\newenvironment{abstractbox}[1][]{

  \ifstrempty{#1}%
  {\mdfsetup{%
    frametitle={%
       \tikz[baseline=(current bounding box.east),outer sep=0pt]
        \node[abstracttitle, anchor=east,rectangle]
        {\strut~~#1~~\strut};}}
  }%
  {\mdfsetup{%
     frametitle={%
       \tikz[baseline=(current bounding box.east),outer sep=0pt]
        \node[abstracttitle,anchor=east,rectangle]
        {\strut~~#1~~\strut};}}%
   }%
   \mdfsetup{innertopmargin=5pt,linecolor=magmadark,%
             linewidth=0.5pt,topline=true,
             frametitleaboveskip=\dimexpr-\ht\strutbox\relax,}
   \begin{mdframed}[backgroundcolor=magmalight,nobreak=true]\relax%
     \raggedright
}{\end{mdframed}}

\tikzstyle{infotitle} =[draw=darkgrey, fill=lightgrey!50, text=black]

\newenvironment{infobox}[1][]{

  \ifstrempty{#1}%
  {\mdfsetup{%
    frametitle={%
       \tikz[baseline=(current bounding box.east),outer sep=0pt]
        \node[infotitle, anchor=east,rectangle]
        {\strut~~#1~~\strut};}}
  }%
  {\mdfsetup{%
     frametitle={%
       \tikz[baseline=(current bounding box.east),outer sep=0pt]
        \node[infotitle,anchor=east,rectangle]
        {\strut~~#1~~\strut};}}%
   }%
   \mdfsetup{innertopmargin=5pt,linecolor=grey,%
             linewidth=0.5pt,topline=true,
             frametitleaboveskip=\dimexpr-\ht\strutbox\relax,}
   \begin{mdframed}[backgroundcolor=lightgrey!25,nobreak=true]\relax%
     \raggedright
}{\end{mdframed}}

\tikzstyle{essencetitle} = [draw=magmadark!75, fill=magmamedium!75, text=black]

\usepackage{enumitem}

\tikzstyle{changelogtitle} =[draw=darkgrey, fill=lightgrey!50, text=black]

\usepackage{listings}

\usepackage{stringstrings}
\usepackage{readarray}

\def\firstchar#1#2|{#1}
\edef\tbs{\detokenize{\X}}
\edef\tbs{\expandafter\firstchar\tbs|}
\edef\tlb{\detokenize{{}}}
\edef\tlb{\expandafter\firstchar\tlb|}
\edef\tus{\detokenize{_}}
\newcounter{index}
\newcommand\detokenizeplus[1]{%
  \def\temparg{\detokenize{#1}}%
  \getargsC{\temparg}%
  \setcounter{index}{0}%
  \def\prevmacro{F}%
  \whiledo{\value{index} < \narg}{%
    \stepcounter{index}%
    \isnextbyte[q]{\tbs}{\csname arg\roman{index}\endcsname}%
    \if T\theresult%
      \if T\prevmacro\unskip\else\fi%
      \def\prevmacro{T}%
    \else%
      \def\prevmacro{F}%
   \fi%
    \isnextbyte[q]{\tlb}{\csname arg\roman{index}\endcsname}%
    \if T\theresult\unskip\else\fi%
    \isnextbyte[q]{\tus}{\csname arg\roman{index}\endcsname}%
    \if T\theresult\unskip\else\fi%
    \csname arg\roman{index}\endcsname~%
  }%
}

\makeatletter
\newread\pin@file
\newcounter{pinlineno}
\newcommand\pin@accu{}
\newcommand\pin@ext{pintmp}
\newcommand*\partialinput [3] {%
  \IfFileExists{#3}{%
    \openin\pin@file #3
    \setcounter{pinlineno}{1}
    \@whilenum\value{pinlineno}<#1 \do{%
      \read\pin@file to\pin@line
      \stepcounter{pinlineno}%
    }
    \addtocounter{pinlineno}{-1}
    \let\pin@accu\empty
    \begingroup
    \endlinechar\newlinechar
    \@whilenum\value{pinlineno}<#2 \do{%
      \readline\pin@file to\pin@line
      \edef\pin@accu{\pin@accu\pin@line}%
      \stepcounter{pinlineno}%
    }
    \closein\pin@file
    \expandafter\endgroup
    \scantokens\expandafter{\pin@accu}%
  }{%
    \errmessage{File `#3' doesn't exist!}%
  }%
}
\makeatother

\usepackage{varioref}

\usepackage{graphicx}
\usepackage{epsfig}
\usepackage{verbatim}

\usepackage{enumerate}
\usepackage{enumitem}

\usepackage{amssymb}
\usepackage{amsmath}

\usepackage{ifthen}

\usepackage{longtable}

\usepackage{mathtools}

\newboolean{twocolswitch}

\usepackage{stackengine}

\usepackage{array}

\usepackage[export]{adjustbox}

\newcommand{\PreserveBackslash}[1]{\let\temp=\\#1\let\\=\temp}

\newcolumntype{P}[1]{>{\centering\arraybackslash}p{#1}}
\newcolumntype{M}[1]{>{\centering\arraybackslash}m{#1}}

\makeatletter
\newcommand*{\centerfloat}{%
  \parindent \z@
  \leftskip \z@ \@plus 1fil \@minus \textwidth
  \rightskip\leftskip
  \parfillskip \z@skip}
\makeatother

\newcommand{\sindex}[1]{}
\newcommand{\nindex}[1]{}

\newcommand{\www}[1]{\url{#1}}

\usepackage{lettrine}

\newcommand{\textmatrix}[1]{\mathbf{#1}}

\newcommand{\lmat}{\left[
    \begin{array}
    }
    \newcommand{\rmat}{\end{array}
  \right]
}

\newcommand{\semdiffsign}{\Leftrightarrow}
\newcommand{\semdiffsignleft}{\Leftarrow}
\newcommand{\semdiffsignright}{\Rightarrow}

\newcommand{\semdiff}[2]{\{#1\,$\semdiffsign$\,#2\}}

\newcommand{\semdiffright}[2]{\{#1\,$\semdiffsignright$\,#2\}}

\newcommand{\semdiffbold}[2]{\{\textbf{#1}\,$\semdiffsign$\,\textbf{#2}\}}
\newcommand{\semdiffboldleft}[2]{\{\textbf{#1}\,$\semdiffsignleft$\,#2\}}
\newcommand{\semdiffboldright}[2]{\{#1\,$\semdiffsignright$\,\textbf{#2}\}}

\newcommand{\semdiffmath}[2]{\{\textnormal{\textbf{#1}}\!\semdiffsign\!{\textnormal{\textbf{#2}}\}}}
\newcommand{\semdiffmathleft}[2]{\{\textnormal{\textbf{#1}}\!\semdiffsign\!{\textnormal{#2}\}}}
\newcommand{\semdiffmathright}[2]{\{\textnormal{#1}\!\semdiffsign\!{\textnormal{\textbf{#2}}\}}}

\newcommand{\archetype}[1]{\archetypelinkbase{#1}}

\newcommand{\datasetsymbol}{\mathbf{D}}
\newcommand{\dataset}[1]{\datasetsymbol_{#1}}

\newcommand{\datasetbase}[1]{
    \IfEqCase{#1}{
        {0800}{\datasetsymbol_{1}}
        {1600}{\datasetsymbol_{2}}
        {2000}{\datasetsymbol_{3}}
    }[\PackageError{datasetbase}{Undefined option to datasetbase: #1}{}]%
}%

\newcommand{\datasetNcharacters}[1]{
    \IfEqCase{#1}{
        {1}{800}
        {2}{1600}
        {3}{2000}
    }[\PackageError{datasetNcharacters}{Undefined option to datasetNcharacters: #1}{}]%
}%

\newcommand{\datasetNtraits}[1]{
    \IfEqCase{#1}{
        {1}{235}
        {2}{364}
        {3}{464}
    }[\PackageError{datasetNtraits}{Undefined option to datasetNtraits: #1}{}]%
}%

\newcommand{\datasetNstories}[1]{
    \IfEqCase{#1}{
        {1}{90}
        {2}{241}
        {3}{341}
    }[\PackageError{datasetNstories}{Undefined option to datasetNstories: #1}{}]%
}%

\newcommand{\Ntraits}{464}
\newcommand{\Ncharacters}{2000}
\newcommand{\Nstories}{341}

\newcommand{\padzero}[1]{\ifnum #1 < 10 0\fi #1}

\usepackage{siunitx}

\newcommand\zeropad[2]{%
  \ifnum#2<0\relax%
    {\ensuremath-}\zeropadA{#1}{\the\numexpr#2*-1\relax}%
  \else%
    \zeropadA{#1}{#2}%
  \fi%
}
\def\zeropadA#1#2{%
  \ifnum1#2<1#1
    \zeropadA{#1}{0#2}%
  \else%
    #2%
  \fi%
}

\usepackage{xstring}

\newcommand{\archetypesemdiff}[1]{
    \IfEqCase{#1}{
        {1}{\semdiffbold{\archetypelinkbase{Fool}}{\archetypelinkbase{Hero}}}                  
        {2}{\semdiffbold{\archetypelinkbase{Angel}}{\archetypelinkbase{Demon}}}                
        {3}{\semdiffbold{\archetypelinkbase{Traditionalist}}{\archetypelinkbase{Adventurer}}}  
        {4}{\semdiffbold{\archetypelinksimple{Lone-Wolf}{Lone Wolf}}{\archetypelinkbase{Diva}}}
        {5}{\semdiffbold{\archetypelinkbase{Outcast}}{\archetypelinkbase{Sophisticate}}}       
        {6}{\semdiffbold{\archetypelinkbase{Brute}}{\archetypelinkbase{Geek}}}                 
    }[\PackageError{archetypesemdiff}{Undefined option to archetypesemdiff: #1}{}]%
}%

\newcommand{\archetypesemdiffleft}[1]{
    \IfEqCase{#1}{
        {1}{\semdiffboldleft{\archetypelinkbase{Fool}}{\archetypelinkbase{Hero}}}                  
        {2}{\semdiffboldleft{\archetypelinkbase{Angel}}{\archetypelinkbase{Demon}}}                
        {3}{\semdiffboldleft{\archetypelinkbase{Traditionalist}}{\archetypelinkbase{Adventurer}}}  
        {4}{\semdiffboldleft{\archetypelinksimple{Lone-Wolf}{Lone Wolf}}{\archetypelinkbase{Diva}}}
        {5}{\semdiffboldleft{\archetypelinkbase{Outcast}}{\archetypelinkbase{Sophisticate}}}       
        {6}{\semdiffboldleft{\archetypelinkbase{Brute}}{\archetypelinkbase{Geek}}}                 
    }[\PackageError{archetypesemdiff}{Undefined option to archetypesemdiff: #1}{}]%
}%

\newcommand{\archetypesemdiffright}[1]{
    \IfEqCase{#1}{
        {1}{\semdiffboldright{\archetypelinkbase{Fool}}{\archetypelinkbase{Hero}}}                  
        {2}{\semdiffboldright{\archetypelinkbase{Angel}}{\archetypelinkbase{Demon}}}                
        {3}{\semdiffboldright{\archetypelinkbase{Traditionalist}}{\archetypelinkbase{Adventurer}}}  
        {4}{\semdiffboldright{\archetypelinksimple{Lone-Wolf}{Lone Wolf}}{\archetypelinkbase{Diva}}}
        {5}{\semdiffboldright{\archetypelinkbase{Outcast}}{\archetypelinkbase{Sophisticate}}}       
        {6}{\semdiffboldright{\archetypelinkbase{Brute}}{\archetypelinkbase{Geek}}}                 
    }[\PackageError{archetypesemdiff}{Undefined option to archetypesemdiff: #1}{}]%
}%

\newcommand{\archetypesemdiffmath}[1]{
  \IfEqCase{#1}{
    {1}{\semdiffmath{\archetypelinkbase{Fool}}{\archetypelinkbase{Hero}}}
    {2}{\semdiffmath{\archetypelinkbase{Angel}}{\archetypelinkbase{Demon}}}
    {3}{\semdiffmath{\archetypelinkbase{Traditionalist}}{\archetypelinkbase{Adventurer}}}
    {4}{\semdiffmath{\archetypelinksimple{Lone-Wolf}{Lone Wolf}}{\archetypelinkbase{Diva}}}
    {5}{\semdiffmath{\archetypelinkbase{Outcast}}{\archetypelinkbase{Sophisticate}}}
    {6}{\semdiffmath{\archetypelinkbase{Brute}}{\archetypelinkbase{Geek}}}                 
  }[\PackageError{archetypesemdiffmath}{Undefined option to archetypesemdiffmath: #1}{}]%
}%

\newcommand{\archetypesemdiffmathswap}[1]{
  \IfEqCase{#1}{
    {1}{\semdiffmath{\archetypelinkbase{Hero}}{\archetypelinkbase{Fool}}}
    {2}{\semdiffmath{\archetypelinkbase{Demon}}{\archetypelinkbase{Angel}}}
    {3}{\semdiffmath{\archetypelinkbase{Adventurer}}{\archetypelinkbase{Traditionalist}}}
    {4}{\semdiffmath{\archetypelinkbase{Diva}}{\archetypelinksimple{Lone-Wolf}{Lone Wolf}}}
    {5}{\semdiffmath{\archetypelinkbase{Sophisticate}}{\archetypelinkbase{Outcast}}}
    {6}{\semdiffmath{{\archetypelinkbase{Geek}}\archetypelinkbase{Brute}}}
  }[\PackageError{archetypesemdiffmathswap}{Undefined option to archetypesemdiffmathswap: #1}{}]%
}%

\newcommand{\archetypesemdiffmathleft}[1]{
  \IfEqCase{#1}{
    {1}{\semdiffmathleft{\archetypelinkbase{Fool}}{\archetypelinkbase{Hero}}}                  
    {2}{\semdiffmathleft{\archetypelinkbase{Angel}}{\archetypelinkbase{Demon}}}                
    {3}{\semdiffmathleft{\archetypelinkbase{Traditionalist}}{\archetypelinkbase{Adventurer}}}  
    {4}{\semdiffmathleft{\archetypelinksimple{Lone-Wolf}{Lone Wolf}}{\archetypelinkbase{Diva}}}
    {5}{\semdiffmathleft{\archetypelinkbase{Outcast}}{\archetypelinkbase{Sophisticate}}}       
    {6}{\semdiffmathleft{\archetypelinkbase{Brute}}{\archetypelinkbase{Geek}}}                 
  }[\PackageError{archetypesemdiffmathleft}{Undefined option to archetypesemdiffmathleft: #1}{}]%
}%

\newcommand{\archetypesemdiffmathright}[1]{
  \IfEqCase{#1}{
    {1}{\semdiffmathright{\archetypelinkbase{Fool}}{\archetypelinkbase{Hero}}}                  
    {2}{\semdiffmathright{\archetypelinkbase{Angel}}{\archetypelinkbase{Demon}}}                
    {3}{\semdiffmathright{\archetypelinkbase{Traditionalist}}{\archetypelinkbase{Adventurer}}}  
    {4}{\semdiffmathright{\archetypelinksimple{Lone-Wolf}{Lone Wolf}}{\archetypelinkbase{Diva}}}
    {5}{\semdiffmathright{\archetypelinkbase{Outcast}}{\archetypelinkbase{Sophisticate}}}       
    {6}{\semdiffmathright{\archetypelinkbase{Brute}}{\archetypelinkbase{Geek}}}                 
  }[\PackageError{archetypesemdiffmathright}{Undefined option to archetypesemdiffmathright: #1}{}]%
}%

\newcommand{\essentialsemdiff}[1]{
  \IfEqCase{#1}{
    {1}{\semdiff{\essentialtraitlinknegative{1}{weak/incompetent/lazy/stupid}}{\essentialtraitlinkpositive{1}{powerful/capable/purposeful/intelligent}}}
    {2}{\semdiff{\essentialtraitlinknegative{2}{safe/pure/virtuous/humble}}{\essentialtraitlinkpositive{2}{dangerous/depraved/corrupt/arrogant}}}
    {3}{\semdiff{\essentialtraitlinknegative{3}{serious/predictable/humorless/uncreative}}{\essentialtraitlinkpositive{3}{playful/unpredictable/funny/creative}}}
    {4}{\semdiff{\essentialtraitlinknegative{4}{rugged/stoic/independent/blunt}}{\essentialtraitlinkpositive{4}{refined/dramatic/dependent/sensitive}}}
    {5}{\semdiff{\essentialtraitlinknegative{5}{unlucky/unsophisticated/traumatized}}{\essentialtraitlinkpositive{5}{fortunate/sophisticated/confident}}}
    {6}{\semdiff{\essentialtraitlinknegative{6}{physical/mainstream/simple-minded}}{\essentialtraitlinkpositive{6}{intellectual/weird/complex}}}
    {7}{\semdiff{\essentialtraitlinknegative{7}{dramatic/attractive/young}}{\essentialtraitlinkpositive{7}{comedic/ugly/old}}}
    {8}{\semdiff{\essentialtraitlinknegative{8}{spiritual/rural/historical}}{\essentialtraitlinkpositive{8}{skeptical/urban/modern}}}
    {9}{\semdiff{\essentialtraitlinknegative{9}{old/historical/low-tempo}}{\essentialtraitlinkpositive{9}{young/modern/high-tempo}}}
    {10}{\semdiff{\essentialtraitlinknegative{10}{feminine/luddite}}{\essentialtraitlinkpositive{10}{masculine/technophile}}}
    {11}{\semdiff{\essentialtraitlinknegative{11}{secondary/street-wise}}{\essentialtraitlinkpositive{11}{primary/sheltered}}}
  }[\PackageError{essentialsemdiff}{Undefined option to essentialsemdiff: #1}{}]%
}%

\newcommand{\essentialsemdiffloose}[1]{
  \IfEqCase{#1}{
    {1}{\semdiff{\essentialtraitlinknegative{1}{weak, incompetent, lazy, stupid}}{\essentialtraitlinkpositive{1}{powerful, capable, purposeful, intelligent}}}
    {2}{\semdiff{\essentialtraitlinknegative{2}{safe, pure, virtuous, humble}}{\essentialtraitlinkpositive{2}{dangerous, depraved, corrupt, arrogant}}}
    {3}{\semdiff{\essentialtraitlinknegative{3}{serious, predictable, humorless, uncreative}}{\essentialtraitlinkpositive{3}{playful, unpredictable, funny, creative}}}
    {4}{\semdiff{\essentialtraitlinknegative{4}{rugged, stoic, independent, blunt}}{\essentialtraitlinkpositive{4}{refined, dramatic, dependent, sensitive}}}
    {5}{\semdiff{\essentialtraitlinknegative{5}{unlucky, unsophisticated, traumatized}}{\essentialtraitlinkpositive{5}{fortunate, sophisticated, confident}}}
    {6}{\semdiff{\essentialtraitlinknegative{6}{physical, mainstream, simple-minded}}{\essentialtraitlinkpositive{6}{intellectual, weird, complex}}}
    {7}{\semdiff{\essentialtraitlinknegative{7}{dramatic, attractive, young}}{\essentialtraitlinkpositive{7}{comedic, ugly, old}}}
    {8}{\semdiff{\essentialtraitlinknegative{8}{spiritual, rural, historical}}{\essentialtraitlinkpositive{8}{skeptical, urban, modern}}}
    {9}{\semdiff{\essentialtraitlinknegative{9}{old, historical, low-tempo}}{\essentialtraitlinkpositive{9}{young, modern, high-tempo}}}
    {10}{\semdiff{\essentialtraitlinknegative{10}{feminine, luddite}}{\essentialtraitlinkpositive{10}{masculine, technophile}}}
    {11}{\semdiff{\essentialtraitlinknegative{11}{secondary, street-wise}}{\essentialtraitlinkpositive{11}{primary, sheltered}}}
  }[\PackageError{essentialsemdiffloose}{Undefined option to essentialsemdiffloose: #1}{}]%
}%

\newcommand{\essentialsemdifflooseleft}[1]{
    \IfEqCase{#1}{
        {1}{\semdiff{\textbf{\essentialtraitlinknegative{1}{weak, incompetent, lazy, stupid}}}{\essentialtraitlinkpositive{1}{powerful, capable, purposeful, intelligent}}}
        {2}{\semdiff{\textbf{\essentialtraitlinknegative{2}{safe, pure, virtuous, humble}}}{\essentialtraitlinkpositive{2}{dangerous, depraved, corrupt, arrogant}}}
        {3}{\semdiff{\textbf{\essentialtraitlinknegative{3}{serious, predictable, humorless, uncreative}}}{\essentialtraitlinkpositive{3}{playful, unpredictable, funny, creative}}}
        {4}{\semdiff{\textbf{\essentialtraitlinknegative{4}{rugged, stoic, independent, blunt}}}{\essentialtraitlinkpositive{4}{refined, dramatic, dependent, sensitive}}}
        {5}{\semdiff{\textbf{\essentialtraitlinknegative{5}{unlucky, unsophisticated, traumatized}}}{\essentialtraitlinkpositive{5}{fortunate, sophisticated, confident}}}
        {6}{\semdiff{\textbf{\essentialtraitlinknegative{6}{physical, mainstream, simple-minded}}}{\essentialtraitlinkpositive{6}{intellectual, weird, complex}}}
        {7}{\semdiff{\textbf{\essentialtraitlinknegative{7}{dramatic, attractive, young}}}{\essentialtraitlinkpositive{7}{comedic, ugly, old}}}
        {8}{\semdiff{\textbf{\essentialtraitlinknegative{8}{spiritual, rural, historical}}}{\essentialtraitlinkpositive{8}{skeptical, urban, modern}}}
        {9}{\semdiff{\textbf{\essentialtraitlinknegative{9}{old, historical, low-tempo}}}{\essentialtraitlinkpositive{9}{young, modern, high-tempo}}}
        {10}{\semdiff{\textbf{\essentialtraitlinknegative{10}{feminine, luddite}}}{\essentialtraitlinkpositive{10}{masculine, technophile}}}
        {11}{\semdiff{\textbf{\essentialtraitlinknegative{11}{secondary, street-wise}}}{\essentialtraitlinkpositive{11}{primary, sheltered}}}
    }[\PackageError{essentialsemdifflooseleft}{Undefined option to essentialsemdifflooseleft: #1}{}]%
 }%

\newcommand{\essentialsemdifflooseright}[1]{
  \IfEqCase{#1}{
    {1}{\semdiff{\essentialtraitlinknegative{1}{weak, incompetent, lazy, stupid}}{\textbf{\essentialtraitlinkpositive{1}{powerful, capable, purposeful, intelligent}}}}
    {2}{\semdiff{\essentialtraitlinknegative{2}{safe, pure, virtuous, humble}}{\textbf{\essentialtraitlinkpositive{2}{dangerous, depraved, corrupt, arrogant}}}}
    {3}{\semdiff{\essentialtraitlinknegative{3}{serious, predictable, humorless, uncreative}}{\textbf{\essentialtraitlinkpositive{3}{playful, unpredictable, funny, creative}}}}
    {4}{\semdiff{\essentialtraitlinknegative{4}{rugged, stoic, independent, blunt}}{\textbf{\essentialtraitlinkpositive{4}{refined, dramatic, dependent, sensitive}}}}
    {5}{\semdiff{\essentialtraitlinknegative{5}{unlucky, unsophisticated, traumatized}}{\textbf{\essentialtraitlinkpositive{5}{fortunate, sophisticated, confident}}}}
    {6}{\semdiff{\essentialtraitlinknegative{6}{physical, mainstream, simple-minded}}{\textbf{\essentialtraitlinkpositive{6}{intellectual, weird, complex}}}}
    {7}{\semdiff{\essentialtraitlinknegative{7}{dramatic, attractive, young}}{\textbf{\essentialtraitlinkpositive{7}{comedic, ugly, old}}}}
    {8}{\semdiff{\essentialtraitlinknegative{8}{spiritual, historical, rural}}{\textbf{\essentialtraitlinkpositive{8}{skeptical, urban, modern}}}}
    {9}{\semdiff{\essentialtraitlinknegative{9}{old, historical, low-tempo}}{\textbf{\essentialtraitlinkpositive{9}{young, modern, high-tempo}}}}
    {10}{\semdiff{\essentialtraitlinknegative{10}{feminine, luddite}}{\textbf{\essentialtraitlinkpositive{10}{masculine, technophile}}}}
    {11}{\semdiff{\essentialtraitlinknegative{11}{secondary, street-wise}}{\textbf{\essentialtraitlinkpositive{11}{primary, sheltered}}}}
  }[\PackageError{essentialsemdifflooseright}{Undefined option to essentialsemdifflooseright: #1}{}]%
}%

\newcommand{\essentialsemdiffmathleft}[1]{
    \IfEqCase{#1}{
        {1}{\semdiffmathleft{weak/incompetent/lazy/stupid}{powerful/capable/purposeful/intelligent}}
        {2}{\semdiffmathleft{safe/pure/virtuous/humble}{dangerous/depraved/corrupt/arrogant}}
        {3}{\semdiffmathleft{serious/predictable/humorless/uncreative}{playful/unpredictable/funny/creative}}
        {4}{\semdiffmathleft{rugged/stoic/independent/blunt}{refined/dramatic/dependent/sensitive}}
        {5}{\semdiffmathleft{unlucky/unsophisticated/traumatized}{fortunate/sophisticated/confident}}
        {6}{\semdiffmathleft{physical/mainstream/simple-minded}{intellectual/weird/complex}}
        {7}{\semdiffmathleft{dramatic/attractive/young}{comedic/ugly/old}}
        {8}{\semdiffmathleft{spiritual/rural/historical}{skeptical/urban/modern}}
        {9}{\semdiffmathleft{old/historical/low-tempo}{young/modern/high-tempo}}
        {10}{\semdiffmathleft{feminine/luddite}{masculine/technophile}}
        {11}{\semdiffmathleft{secondary/street-wise}{primary/sheltered}}
    }[\PackageError{essentialsemdiffmathleft}{Undefined option to essentialsemdiffmathleft: #1}{}]%
}%

\newcommand{\essentialsemdiffmathright}[1]{
    \IfEqCase{#1}{
        {1}{\semdiffmathright{weak/incompetent/lazy/stupid}{powerful/capable/purposeful/intelligent}}
        {2}{\semdiffmathright{safe/pure/virtuous/humble}{dangerous/depraved/corrupt/arrogant}}
        {3}{\semdiffmathright{serious/predictable/humorless/uncreative}{playful/unpredictable/funny/creative}}
        {4}{\semdiffmathright{rugged/stoic/independent/blunt}{refined/dramatic/dependent/sensitive}}
        {5}{\semdiffmathright{unlucky/unsophisticated/traumatized}{fortunate/sophisticated/confident}}
        {6}{\semdiffmathright{physical/mainstream/simple-minded}{intellectual/weird/complex}}
        {7}{\semdiffmathright{dramatic/attractive/young}{comedic/ugly/old}}
        {8}{\semdiffmathright{spiritual/rural/historical}{skeptical/urban/modern}}
        {9}{\semdiffmathright{old/historical/low-tempo}{young/modern/high-tempo}}
        {10}{\semdiffmathright{feminine/luddite}{masculine/technophile}}
        {11}{\semdiffmathright{secondary/street-wise}{primary/sheltered}}
    }[\PackageError{essentialsemdiffmathright}{Undefined option to essentialsemdiffmathright: #1}{}]%
}%

\newcommand{\essentialsemdiffmath}[1]{
    \IfEqCase{#1}{
        {1}{\semdiffmath{weak/incompetent/lazy/stupid}{powerful/capable/purposeful/intelligent}}
        {2}{\semdiffmath{safe/pure/virtuous/humble}{dangerous/depraved/corrupt/arrogant}}
        {3}{\semdiffmath{serious/predictable/humorless/uncreative}{playful/unpredictable/funny/creative}}
        {4}{\semdiffmath{rugged/stoic/independent/blunt}{refined/dramatic/dependent/sensitive}}
        {5}{\semdiffmath{unlucky/unsophisticated/traumatized}{fortunate/sophisticated/confident}}
        {6}{\semdiffmath{physical/mainstream/simple-minded}{intellectual/weird/complex}}
        {7}{\semdiffmath{dramatic/attractive/young}{comedic/ugly/old}}
        {8}{\semdiffmath{spiritual/rural/historical}{skeptical/urban/modern}}
        {9}{\semdiffmath{old/historical/low-tempo}{young/modern/high-tempo}}
        {10}{\semdiffmath{feminine/luddite}{masculine/technophile}}
        {11}{\semdiffmath{secondary/street-wise}{primary/sheltered}}
    }[\PackageError{essentialsemdiffmath}{Undefined option to essentialsemdiffmath: #1}{}]%
}%

\newcommand{\ousiometricsemdiff}[1]{
    \IfEqCase{#1}{
        {1}{\semdiffbold{weak}{powerful}}
        {2}{\semdiffbold{safe}{dangerous}}
        {3}{\semdiffbold{structured}{unstructured}}
    }[\PackageError{ousiometricsemdiff}{Undefined option to ousiometricsemdiff: #1}{}]%
}%

\newcommand{\ousiometricsemdiffmath}[1]{
    \IfEqCase{#1}{
        {1}{\semdiffmath{weak}{powerful}}
        {2}{\semdiffmath{safe}{dangerous}}
        {3}{\semdiffmath{structured}{unstructured}}
    }[\PackageError{ousiometricsemdiffmath}{Undefined option to ousiometricsemdiffmath: #1}{}]%
}%

\newcommand{\ousiometricsemdiffmathleft}[1]{
    \IfEqCase{#1}{
        {1}{\semdiffmathleft{weak}{powerful}}
        {2}{\semdiffmathleft{safe}{dangerous}}
        {3}{\semdiffmathleft{structured}{unstructured}}
    }[\PackageError{ousiometricsemdiffmathleft}{Undefined option to ousiometricsemdiffmathleft: #1}{}]%
}%

\newcommand{\ousiometricsemdiffmathright}[1]{
    \IfEqCase{#1}{
        {1}{\semdiffmathright{weak}{powerful}}
        {2}{\semdiffmathright{safe}{dangerous}}
        {3}{\semdiffmathright{structured}{unstructured}}
    }[\PackageError{ousiometricsemdiffmathright}{Undefined option to ousiometricsemdiffmathright: #1}{}]%
}%

\newcommand{\dimensiontype}[1]{
    \IfEqCase{#1}{
        {1}{Major archetype}
        {2}{Major Archetype}
        {3}{Major Archetype}
        {4}{Minor Archetype}
        {5}{Minor Archetype}
        {6}{Minor Archetype}
        {7}{Trait}
        {8}{Trait}
        {9}{Trait}
        {10}{Trait}
        {11}{Trait}
    }[\PackageError{dimensiontype}{Undefined option to dimensiontype: #1}{}]%
}%

\newcommand{\traittable}[1]{

  \begin{table*}[thp]
    \centering
    \small
    {
      \large Essential Trait Dimension #1, $\hat{v}_{#1}$

      \ifthenelse{#1 < 7}
      {
      \smallskip
      
      \dimensiontype{#1} dimension: \archetypesemdiff{#1}

      \smallskip

      \essentialsemdiff{#1}
      }{
      \smallskip
      
      \dimensiontype{#1} dimension, non-archetype

      \smallskip

      \essentialsemdiff{#1}
      }
    }
    
    \bigskip

    \rowcolors{3}{traitrowcolor}{traitrowcoloralt}
    \begin{tabular}{>{\raggedright\arraybackslash}p{150pt}ccccc}
      \hline
      \rowcolor{headerpop}
      Most aligned traits ($\hat{v}_{#1}$)
      &
      Cos.
      &
      Var.
      &
      Comp.
      &
      Trait
      &
      Size
      \\
      \rowcolor{headerpop}
      &
      &
      Expl.
      &
      Size
      &
      Size
      &
      Rank
      \\
      \hline
      \input{N\Ncharactersbase_aligned_essential_traits\padzero{#1}.tex}\unskip
      \\ 
    \end{tabular}

    \bigskip
    
    \rowcolors{2}{traitrowcolor}{traitrowcoloralt}
    \begin{tabular}{>{\raggedright\arraybackslash}p{150pt}ccccc}
      \hline
      \rowcolor{headerpop}
      Traits with largest component ($\hat{v}_{#1}$) 
      &
      Cos.
      &
      Var.
      &
      Comp.
      &
      Trait
      &
      Size
      \\
      \rowcolor{headerpop}
      &
      &
      Expl.
      &
      Size
      &
      Size
      &
      Rank
      \\
      \hline
      \input{N\Ncharactersbase_aligned_essential_traits_component_size\padzero{#1}.tex}\unskip
      \\ 
    \end{tabular}
    
    \caption{
      Dominant traits
      making up essential dimension #1, \!\!\!\!\protect\ifthenelse{#1 < 7}
      {
      the \protect\dimensiontype{#1} dimension \!\!\protect\archetypesemdiff{#1}
      }{
      a \protect\dimensiontype{#1} dimension (non-archetype)
      }
      which is described by the complex semantic differential \protect\essentialsemdiffloose{#1}.
      The first table shows traits ordered by alignment
      with the singular unit vector $\hat{v}_{#1}$,
      and the second re-ordered by component size
      ($\textmatrix{A}\hat{v}_{#1}$).
      As traits are directed semantic differentials, they are flipped
      to be positively aligned, if needed.
      In the trailing parentheses, the numbers
      for each trait are:
      1. Alignment strength (cosine of angle between
      measured semantic differential vectors and
      the unit singular vector $\hat{v}_{#1}$);
      2. Percentage of variance explained;
      and
      3. 
      Component size in the direction of the unit singular vector $\hat{v}_{#1}$;
      4.
      Trait strength as vector magnitude in character space;
      and
      5.
      Trait strength rank out of \Ntraits\ traits.
    }
    \label{tab:archetypometricssupp.N\Ncharactersbase_singular-dimension-trait#1}
  \end{table*}

  \clearpage
}

\newcommand{\charactertable}[1]{

  \begin{table*}[thp]
    \centering
    \small
    {
      \large Essential Character Dimension #1, $\hat{u}_{#1}$

      \ifthenelse{#1 < 7}
      {
      \smallskip
      
      \dimensiontype{#1} dimension: \archetypesemdiff{#1}

      \smallskip

      \essentialsemdiff{#1}
      }{
      \smallskip
      
      \dimensiontype{#1} dimension, non-archetype

      \smallskip

      \essentialsemdiff{#1}
      }
    }

    \bigskip

    \hspace*{-20pt}
    \adjustbox{valign=t}{\rowcolors{2}{characterrowcolor}{characterrowcoloralt}
    \begin{tabular}{>{\raggedright\arraybackslash}p{160pt}ccccc}
      \hline
      \rowcolor{headerpop}
      Most negatively aligned 
      &
      Cos.
      &
      Var.
      &
      Comp.
      &
      Char.
      &
      Size
      \\
      \rowcolor{headerpop}
      characters ($-\hat{u}_{#1}$)
      &
      &
      Expl.
      &
      Size
      &
      Size
      &
      Rank
      \\
      \hline
      
      \input{N\Ncharactersbase_aligned_negative_essential_characters\padzero{#1}.tex}\unskip
      \\
    \end{tabular}}%
    ~~~~%
    \adjustbox{valign=t}{\rowcolors{2}{characterrowcolor}{characterrowcoloralt}%
    \begin{tabular}{>{\raggedright\arraybackslash}p{160pt}ccccc}
      \hline
      \rowcolor{headerpop}
      Most positively aligned 
      &
      Cos.
      &
      Var.
      &
      Comp.
      &
      Char.
      &
      Size
      \\
      \rowcolor{headerpop}
      characters ($+\hat{u}_{#1}$)
      &
      &
      Expl.
      &
      Size
      &
      Size
      &
      Rank
      \\
      \hline
      \input{N\Ncharactersbase_aligned_positive_essential_characters\padzero{#1}.tex}\unskip
      \\
    \end{tabular}}

    \bigskip

    \hspace*{-20pt}
    \adjustbox{valign=t}{\rowcolors{2}{characterrowcolor}{characterrowcoloralt}
    \begin{tabular}{>{\raggedright\arraybackslash}p{160pt}ccccc}
      \hline
      \rowcolor{headerpop}
      Characters with largest 
      &
      Cos.
      &
      Var.
      &
      Comp.
      &
      Char.
      &
      Size
      \\
      \rowcolor{headerpop}
      negative component ($-\hat{u}_{#1}$)
      &
      &
      Expl.
      &
      Size
      &
      Size
      &
      Rank
      \\
      \hline
      \input{N\Ncharactersbase_aligned_negative_essential_characters_component_size\padzero{#1}.tex}\unskip
      \\
    \end{tabular}}%
    ~~~~%
    \adjustbox{valign=t}{\rowcolors{2}{characterrowcolor}{characterrowcoloralt}%
    \begin{tabular}{>{\raggedright\arraybackslash}p{160pt}ccccc}
      \rowcolor{headerpop}
      \hline
      \rowcolor{headerpop}
      Characters with largest 
      &
      Cos.
      &
      Var.
      &
      Comp.
      &
      Char.
      &
      Size
      \\
      \rowcolor{headerpop}
      positive component ($+\hat{u}_{#1}$)
      &
      &
      Expl.
      &
      Size
      &
      Size
      &
      Rank
      \\
      \hline
      \input{N\Ncharactersbase_aligned_positive_essential_characters_component_size\padzero{#1}.tex}\unskip
      \\
    \end{tabular}}

  \caption{
      Leading characters
      for essential dimension #1, \protect\ifthenelse{#1 < 7}
      {
      the \protect\dimensiontype{#1} dimension \!\!\protect\archetypesemdiff{#1}\unskip
      }{
      a \protect\dimensiontype{#1} dimension (non-archetype)\unskip
      }
      which is described by the complex semantic differential \protect\essentialsemdiffloose{#1}.
      In the top two tables,
      characters are arranged by alignment,
      and in the bottom two tables, by strength of component ($\hat{u}_{#1}^{\textnormal{T}}\textmatrix{A}$).
      Tables on the left list negatively aligned characters,
      and tables on the right list positively aligned ones.
      The five numerical columns record:
      1. Alignment strength (cosine of angle between
      individual vectors and the unit singular vector $\hat{u}_{#1}$);
      2. Percentage of variance explained;
      and
      3. 
      Component size in the direction of the unit singular vector $\hat{u}_{#1}$;
      4. 
      Character size as vector magnitude in trait space;
      and
      5.
      Character size rank out of \Ncharacters\ characters.
      See Sec.~\ref{subsec:archetypometricssupp.abbreviations} for
      story abbreviations.
    }
    \label{tab:archetypometricssupp.N\Ncharactersbase_singular-dimension-character#1}
  \end{table*}

  \clearpage
}

\setstackgap{S}{1pt}
\newcommand{\babaisyouboxscale}{0.48}

\newcommand{\archetyperatiosymbol}[2]{{\colorbox{#1}{\textcolor{#2}{\stackanchor{\scalebox{\babaisyouboxscale}{AR}}{\scalebox{\babaisyouboxscale}{CH}}}}}}

\newcommand{\appendixsymbol}[2]{{\colorbox{#1}{\textcolor{#2}{\stackanchor{\scalebox{\babaisyouboxscale}{AP}}{\scalebox{\babaisyouboxscale}{DX}}}}}}

\newcommand{\archetypometricshome}{\onlinesite}
\newcommand{\cardsdir}{\archetypometricshome/cards}

\newcommand{\characterlinksimple}[2]{\href{\cardsdir/#1-\Ncharacters-\Ntraits-\Nstories.pdf}{\textcolor{verydarkgrey}{#2\paperlinksymbol}}}

\newcommand{\characterlinksimpledataset}[3]{
  \IfEqCase{#3}{
    {1}{\href{\cardsdir/#1-\Ncharactersmainone-\Ntraitsmainone-\Nstoriesmainone.pdf}{\textcolor{verydarkgrey}{#2\colorbox{datasetrowcolor}{$\dataset{#3}$}\paperlinksymbol}}}
    {2}{\href{\cardsdir/#1-\Ncharactersmaintwo-\Ntraitsmaintwo-\Nstoriesmaintwo.pdf}{\textcolor{verydarkgrey}{#2\colorbox{datasetrowcolor}{$\dataset{#3}$}\paperlinksymbol}}}
    {3}{\href{\cardsdir/#1-\Ncharactersmain-\Ntraitsmain-\Nstoriesmain.pdf}{\textcolor{verydarkgrey}{#2\colorbox{datasetrowcolor}{$\dataset{#3}$}\paperlinksymbol}}}
  }[\PackageError{characterlinksimpledataset}{Undefined option to characterlinksimpledataset: #1}{}]%
}

\newcommand{\traitlinkonelabel}[3]{\textcolor{verydarkgrey}{\ifthenelse{\equal{#3}{#1}}{\href{\cardsdir/#2--#1-\Ncharacters-\Ntraits-\Nstories.pdf}{#3\paperlinksymbol}}{\href{\cardsdir/#1--#2-\Ncharacters-\Ntraits-\Nstories.pdf}{#3\paperlinksymbol}}}}

\newcommand{\traitlinksimple}[2]{\textcolor{verydarkgrey}{\semdiff{\href{\cardsdir/#2--#1-\Ncharacters-\Ntraits-\Nstories.pdf}{#1\paperlinksymbol}}{\href{\cardsdir/#1--#2-\Ncharacters-\Ntraits-\Nstories.pdf}{#2\paperlinksymbol}}}}

\newcommand{\traitlinksimpledataset}[3]{
  \IfEqCase{#3}{
    {1}{\textcolor{verydarkgrey}{\semdiff{\href{\cardsdir/#2--#1-\Ncharactersmainone-\Ntraitsmainone-\Nstoriesmainone.pdf}{#1\colorbox{datasetrowcolor}{$\dataset{#3}$}\paperlinksymbol}}{\href{\cardsdir/#1--#2-\Ncharactersmainone-\Ntraitsmainone-\Nstoriesmainone.pdf}{#2\colorbox{datasetrowcolor}{$\dataset{#3}$}\paperlinksymbol}}}}
    {2}{\textcolor{verydarkgrey}{\semdiff{\href{\cardsdir/#2--#1-\Ncharactersmaintwo-\Ntraitsmaintwo-\Nstoriesmaintwo.pdf}{#1\colorbox{datasetrowcolor}{$\dataset{#3}$}\paperlinksymbol}}{\href{\cardsdir/#1--#2-\Ncharactersmaintwo-\Ntraitsmaintwo-\Nstoriesmaintwo.pdf}{#2\colorbox{datasetrowcolor}{$\dataset{#3}$}\paperlinksymbol}}}}
    {3}{\textcolor{verydarkgrey}{\semdiff{\href{\cardsdir/#2--#1-\Ncharactersmain-\Ntraitsmain-\Nstoriesmain.pdf}{#1\colorbox{datasetrowcolor}{$\dataset{#3}$}\paperlinksymbol}}{\href{\cardsdir/#1--#2-\Ncharactersmain-\Ntraitsmain-\Nstoriesmain.pdf}{#2\colorbox{datasetrowcolor}{$\dataset{#3}$}\paperlinksymbol}}}}
  }[\PackageError{traitlinksimpledataset}{Undefined option to traitlinksimpledataset: #1}{}]%
}

\newcommand{\traitlinksimpleright}[2]{\href{\cardsdir/#1--#2-\Ncharacters-\Ntraits-\Nstories.pdf}{\textcolor{verydarkgrey}{\semdiffright{#1}{\textbf{#2}}\paperlinksymbol}}}

\newcommand{\traitlinksimplerightalt}[4]{\href{\cardsdir/#1--#2-\Ncharacters-\Ntraits-\Nstories.pdf}{\textcolor{verydarkgrey}{\semdiffright{#3}{\textbf{#4}}\paperlinksymbol}}}

\newcommand{\traitlinksimpledatasetalt}[5]{
  \IfEqCase{#5}{
    {1}{\textcolor{verydarkgrey}{\semdiff{\href{\cardsdir/#2--#1-\Ncharactersmainone-\Ntraitsmainone-\Nstoriesmainone.pdf}{#3\colorbox{datasetrowcolor}{$\dataset{#5}$}\paperlinksymbol}}{\href{\cardsdir/#1--#2-\Ncharactersmainone-\Ntraitsmainone-\Nstoriesmainone.pdf}{#4\colorbox{datasetrowcolor}{$\dataset{#5}$}\paperlinksymbol}}}}
    {2}{\textcolor{verydarkgrey}{\semdiff{\href{\cardsdir/#2--#1-\Ncharactersmaintwo-\Ntraitsmaintwo-\Nstoriesmaintwo.pdf}{#3\colorbox{datasetrowcolor}{$\dataset{#5}$}\paperlinksymbol}}{\href{\cardsdir/#1--#2-\Ncharactersmaintwo-\Ntraitsmaintwo-\Nstoriesmaintwo.pdf}{#4\colorbox{datasetrowcolor}{$\dataset{#5}$}\paperlinksymbol}}}}
    {3}{\textcolor{verydarkgrey}{\semdiff{\href{\cardsdir/#2--#1-\Ncharactersmain-\Ntraitsmain-\Nstoriesmain.pdf}{#3\colorbox{datasetrowcolor}{$\dataset{#5}$}\paperlinksymbol}}{\href{\cardsdir/#1--#2-\Ncharactersmain-\Ntraitsmain-\Nstoriesmain.pdf}{#4\colorbox{datasetrowcolor}{$\dataset{#5}$}\paperlinksymbol}}}}
  }[\PackageError{traitlinksimpledatasetalt}{Undefined option to traitlinksimpledatasetalt: #1}{}]%
}

\newcommand{\storylinksimple}[2]{\href{\cardsdir/#1-\Ncharacters-\Ntraits-\Nstories.pdf}{\textcolor{darkgrey}{#2\paperlinksymbol}}}

\newcommand{\storylinksimpledataset}[3]{
  \IfEqCase{#3}{
    {1}{\href{\cardsdir/#1-\Ncharactersmainone-\Ntraitsmainone-\Nstoriesmainone.pdf}{\textcolor{verydarkgrey}{#2\colorbox{datasetrowcolor}{$\dataset{#3}$}\paperlinksymbol}}}
    {2}{\href{\cardsdir/#1-\Ncharactersmaintwo-\Ntraitsmaintwo-\Nstoriesmaintwo.pdf}{\textcolor{verydarkgrey}{#2\colorbox{datasetrowcolor}{$\dataset{#3}$}\paperlinksymbol}}}
    {3}{\href{\cardsdir/#1-\Ncharactersmain-\Ntraitsmain-\Nstoriesmain.pdf}{\textcolor{verydarkgrey}{#2\colorbox{datasetrowcolor}{$\dataset{#3}$}\paperlinksymbol}}}
  }[\PackageError{storylinksimpledataset}{Undefined option to storylinksimpledataset: #1}{}]%
}

\newcommand{\archetypelinkbase}[1]{\href{\cardsdir/Archetype-#1-component-size-\Ncharacters-\Ntraits-\Nstories.pdf}{\textcolor{verydarkgrey}{#1\paperlinksymbol}}}

\newcommand{\archetypelinksimple}[2]{\href{\cardsdir/Archetype-#1-component-size-\Ncharacters-\Ntraits-\Nstories.pdf}{\textcolor{verydarkgrey}{#2\paperlinksymbol}}}

\newcommand{\archetypelinksimpledataset}[3]{
  \IfEqCase{#3}{
    {1}{\href{\cardsdir/Archetype-#1-component-size-\Ncharactersmainone-\Ntraitsmainone-\Nstoriesmainone.pdf}{\textcolor{verydarkgrey}{#2\colorbox{datasetrowcolor}{$\dataset{#3}$}\paperlinksymbol}}}
    {2}{\href{\cardsdir/Archetype-#1-component-size-\Ncharactersmaintwo-\Ntraitsmaintwo-\Nstoriesmaintwo.pdf}{\textcolor{verydarkgrey}{#2\colorbox{datasetrowcolor}{$\dataset{#3}$}\paperlinksymbol}}}
    {3}{\href{\cardsdir/Archetype-#1-component-size-\Ncharactersmain-\Ntraitsmain-\Nstoriesmain.pdf}{\textcolor{verydarkgrey}{#2\colorbox{datasetrowcolor}{$\dataset{#3}$}\paperlinksymbol}}}
  }[\PackageError{archetypelinksimpledataset}{Undefined option to archetypelinksimpledataset: #1}{}]%
}

\newcommand{\archetypelinkratiosimpleappendix}[2]{{\hypersetup{allcolors=.}\hyperref[page:N\Ncharactersbase_archetypometrics.archetypeclass-#1]{#2\archetyperatiosymbol{archetyperowcolor}{black}\,\appendixsymbol{verydarkgrey}{white}\internallinksymbol}}}

\newcommand{\essentialtraitlinknegative}[2]{\href{\cardsdir/Essential-Trait-\zeropad{000}{#1}-negative-component-size-\Ncharacters-\Ntraits-\Nstories.pdf}{\textcolor{verydarkgrey}{#2\paperlinksymbol}}}

\newcommand{\essentialtraitlinkpositive}[2]{\href{\cardsdir/Essential-Trait-\zeropad{000}{#1}-positive-component-size-\Ncharacters-\Ntraits-\Nstories.pdf}{\textcolor{verydarkgrey}{#2\paperlinksymbol}}}

\newcommand{\characterlink}[1]{
  \IfEqCase{#1}{
    }[\PackageError{characterlink}{Undefined option to characterlink: #1}{}]%
}%

\newcommand{\characterlinkinsert}[2]{
  \IfEqCase{#1}{
    }[\PackageError{characterlinkinsert}{Undefined option to characterlinkinsert: #1}{}]%
}%

\setboolean{twocolswitch}{true}

\usepackage{natbib}
\setcitestyle{square}
\raggedright

\newcommand{\Ncharactersbase}{2000}

\newcommand{\Ncharactersmain}{2000}
\newcommand{\Ntraitsmain}{464}
\newcommand{\Nstoriesmain}{341}

\newcommand{\Ncharactersmainone}{0800}
\newcommand{\Ntraitsmainone}{235}
\newcommand{\Nstoriesmainone}{090}

\newcommand{\Ncharactersmaintwo}{1600}
\newcommand{\Ntraitsmaintwo}{364}
\newcommand{\Nstoriesmaintwo}{241}

\renewcommand{\Ncharacters}{2000}
\renewcommand{\Ntraits}{464}
\renewcommand{\Nstories}{341}

\usepackage{pdfpages}
\usepackage{newpax}
\newpaxsetup{usefileattributes=true}
\makeatletter
\def\newpax@auxfile#1{%
  \immediate\openout\newpax@auxout=%
    \expandafter\newpax@auxname\csname newpax@aux@#1\endcsname%
}
\def\newpax@auxname#1{%
  \newwrite\newpax@auxout
  \immediate\openout\newpax@auxout=build/#1.newpax%
  \immediate\write\newpax@auxout{\relax}%
}
\makeatother

\usepackage{authblk}

\begin{document}

\title{\protect
Buffy versus Bella: An archetypometric analysis and comparison

}

\renewcommand*{\Authsep}{, }
\renewcommand*{\Authand}{, }
\renewcommand*{\Authands}{, }
\renewcommand*{\Affilfont}{\normalsize\normalfont}
\renewcommand*{\Authfont}{\bfseries}
\setlength{\affilsep}{2em}

\author[1\thanks{calla.beauregard@uvm.edu}]{Calla~Glavin~Beauregard}
\author[1,2]{Julia~Witte~Zimmerman}
\author[1]{Ashley~M.~A.~Fehr}
\author[3]{Savannah Hinson Rivas}
\author[1]{Tabia Tanzin Prama}
\author[1]{Tessa~Lawler}
\author[4]{Timothy~R.~Tangherlini}
\author[1,5]{Christopher~M.~Danforth}
\author[1,6,7,8\thanks{peter.dodds@uvm.edu}]{Peter~Sheridan~Dodds}

\affil[1]{
  Computational Story Lab,
  Vermont Advanced Computing Center,
  Vermont Complex Systems Institute,
  MassMutual Center of Excellence for Complex Systems and Data Science,
  University of Vermont,
  Burlington,
  VT 05405,
  US
}
\affil[2]{
  Computational Ethics Lab,
  University of Vermont,
  Burlington,
  VT 05405,
  US
}

\affil[3]{
Department of Linguistics,
University~of~New~Mexico,~Albuquerque,~NM,~87131-0001,~USA
}

\affil[4]{
  Department of Scandinavian,
  Folklore Program,
  School of Information,
  Berkeley Institute for Data Science,
  University~of~California,~Berkeley,~Berkeley,~CA~94720-1500,~USA
}

\affil[5]{
  Department of Mathematics \& Statistics,
  University of Vermont,
  Burlington,
  VT 05405,
  US
  }

\affil[6]{
  Department of Computer Science,
  University of Vermont,
  Burlington,
  VT 05405,
  US
}

\affil[7]{
  Santa Fe Institute,
  1399 Hyde Park Rd,
  Santa Fe,
  NM 87501,
  US
}

\affil[8]{
  Complexity Science Hub,
  Metternichgasse 8,
  1030 Vienna,
  Austria
}

\date{\today}


\maketitle



\mbox{}

\bigskip
\bigskip
\bigskip

\hspace*{+80pt}
\begin{minipage}{300pt}

  \begin{tabular}{>{\raggedright\arraybackslash}p{180pt}p{30pt}p{180pt}}

    \parbox{180pt}{    
      \begin{loglinebox}[Logline]
        \raggedright
        We use archetypometrics to 
quantify 
and
compare 
the characters of Buffy Summers and Bella Swan, 
protagonists of widely popular vampire stories.  
        \smallskip
      \end{loglinebox}
    }

    &
    
    &
    
    \parbox{180pt}{    
      \includegraphics[width=\linewidth]{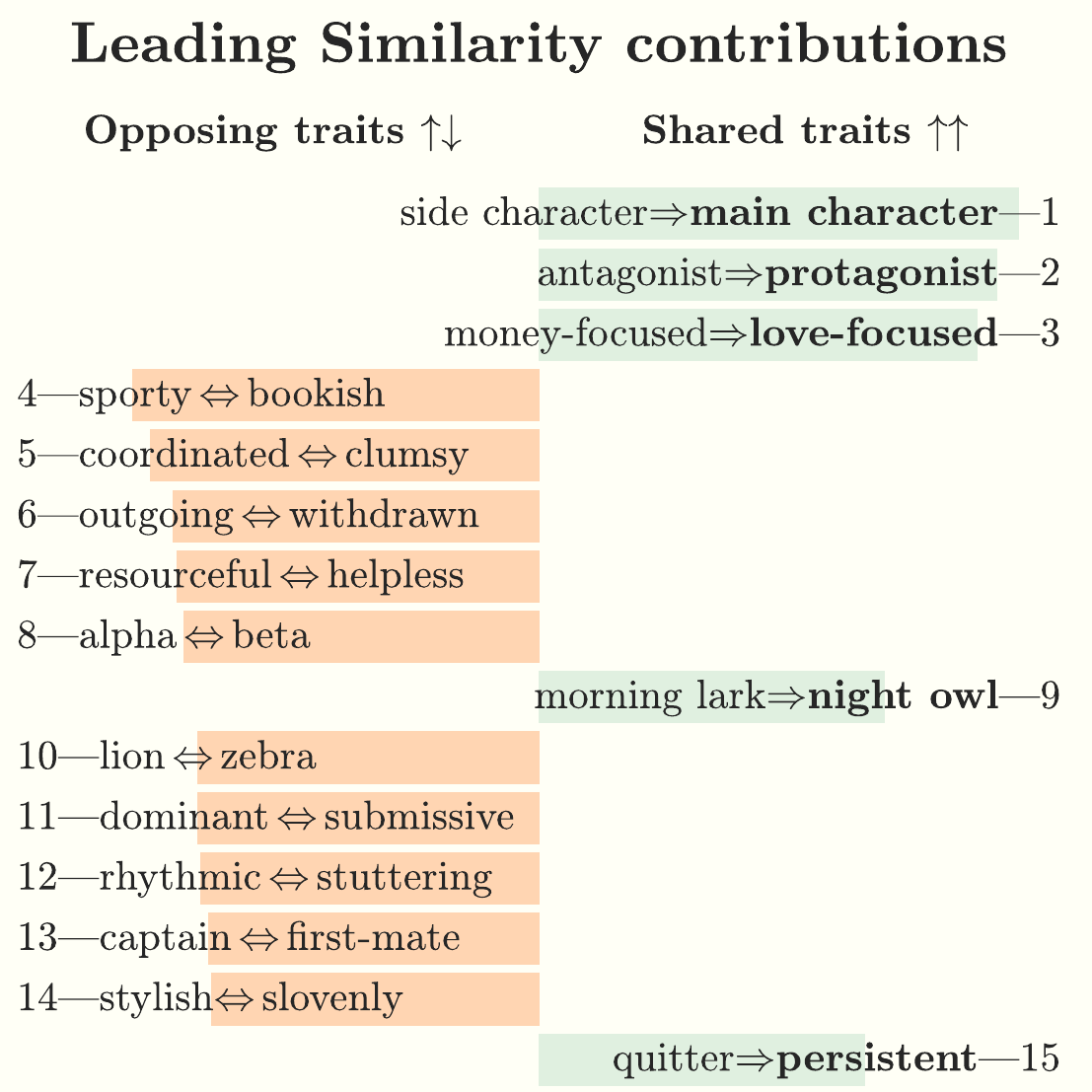}
    }
    
  \end{tabular}
  
\end{minipage}

\clearpage

\newgeometry{
  left=2in,
  right=2in,
  top=1in,
  bottom=1in,
  }

\onecolumn

\renewcommand{\baselinestretch}{1.25}
\selectfont


\begin{abstractbox}[Abstract:]
  \raggedright
  Fictional stories and
characters embody and encode social norms, and their study is a powerful tool through which to understand culture and society. 
Vampire stories and folklore, in particular, have long both reflected and refracted people's 
preoccupation with 
disease,
sexuality,
death,
and
immortality.
Here, we explore female main characters from two popular vampire franchises of the 21st century: Buffy Summers from the eponymous Buffy the Vampire Slayer and Bella Swan from the Twilight series. 
We employ the archetypometrics framework, built from 
2,000 characters assesed across 464 semantic differential traits, 
to understand Buffy's and Bella's archetypes compared to one another and characters in their own stories, as well as within a larger societal context. 
While Buffy and Bella are female protagonists who share focus on love and romance, 
they differ broadly on their underlying traits
and overall archetypes.
Buffy---presented as a prototypical high school cheerleader---largely bucks traditional gender norms as an strong Adventurer-Hero. 
Bella---stylized as ``not like the other girls''---largely conforms to traditional gender norms as a weak Outcast archetype. 
In each instance, our use of archetypometrics offers a detailed, character-based lens for assessing female protagonists in contemporary vampire narratives, 
with clear potential for broader application across other storytelling forms.
  \smallskip
\end{abstractbox}

\vfill


\begin{infobox}[Keywords:]
  \centering
  archetypes,
computational stories,
cultural analysis,
Buffy the Vampire Slayer,
Twilight,
television,
film

 \smallskip
\end{infobox}

\renewcommand{\baselinestretch}{1}
\selectfont

\twocolumn

\restoregeometry

\clearpage

\tableofcontents

\clearpage

\section{Introduction}
\label{sec:buffy-versus-bella.introduction}

Fictional stories, in general, encode and relate fundamental aspects of human societies~\cite{mar2008fiction}. 
Reading fiction can promote personal insight~\cite{oatley1999fiction} and build empathy responses in readers~\cite{bal2013fiction}, and movies can change young people's attitudes towards social issues~\cite{kubrak2020films}. 
Fictional characters embody social norms and their study is a powerful tool through which to understand society~\cite{piper2023characters}. 

In this short paper, we focus in detail on a specific pair of female characters, chosen for their striking contrast: \characterlinksimple{Buffy-the-Vampire-Slayer-Buffy-Summers}{Buffy}, from \storylinksimple{Buffy-the-Vampire-Slayer}{Buffy the Vampire Slayer}~\cite{giannini2023buffy}, and \characterlinksimple{Twilight-Bella-Swan}{Bella}, from the \storylinksimple{Twilight}{the Twilight series}~\cite{larsson2011twilight}. Both are modernized retellings of the vampire's tale, with the focus shifted to the vampire's female counterpart -- traditionally ``victim'' and/or love interest -- as heroine, though focused in different ways. Both stories offer identification with the reader through the heroine's strength, extraordinary capabilities, status as an object of desire, or some combination of these traits~\cite{MatekLukic2013BellaBeast}.
The character of Buffy intentionally subverts older tropes in favor of portraying a feminist, enlightened hero~\cite{pender2004}. 
On the other hand, Bella Swan is something of a ``self-insert'' -- as can be seen even in her name (Bella Swan, Beautiful Swan, the nonspecific evolution of the Ugly Duckling) -- into a very familiar romance (hearkening to Tristan and Iseult, Wuthering Heights, and Romeo and Juliet), with a tension-driving emphasis on chastity, restraint over carnal desire, and marriage~\cite{Kokkola2011VirtuousVampires}. 

Previous scholarly work is mixed on the representation of these stories' respective female protagonists~\cite{nicol2011whenyoukiss,jones2013buffy,coker-2011-bella-buffy-ethics}.
Some scholars laud Buffy as an embodiment of feminist empowerment~\cite{nicol2011whenyoukiss}, whereas others remark that both Buffy or Bella---for different reasons---promote traditional gender ideals~\cite{jones2013buffy}. 
In either case, \characterlinksimple{Buffy-the-Vampire-Slayer-Buffy-Summers}{Buffy} and \characterlinksimple{Twilight-Bella-Swan}{Bella} are interesting characters both in their contrast to each other and their departure from the traditional female counterparts of vampires: innocent, virginal victim, or damned, worldly, seductive lover~\cite{pollard2016,brodman2013,mellins2013}.  
Moreover, 
Buffy the Vampire Slayer 
and 
Twilight
remain widely popular internationally~\cite{giannini2023buffy,larsson2011twilight,stasiewicz-bienkowska2023,mcintosh-2009-buffy-vs-edward} among young people, suggesting sustained cultural impact~\cite{kubrak2020films}. 

Here, we add to the extant scholarship on
\characterlinksimple{Buffy-the-Vampire-Slayer-Buffy-Summers}{Buffy}
and 
\characterlinksimple{Twilight-Bella-Swan}{Bella}
through a robust, quantified spatial representation of character archetypes, which is itself grounded in essential meaning~\cite{dodds2026b}.

In Ref.~\cite{dodds_2025_17128112_pragmateia_archetypes}, 
we have developed archetypometrics: 
A framework to quantitatively describe understand fictional characters in TV, film, and literature 
based on surveyed judgments of semantic differentials
like
\traitlinksimple{quitter}{persistent}
and
\traitlinksimple{kind}{cruel}.
We uncovered six base Archetype pairs, partitioned into three primary and three secondary orthogonal dimensions:
\archetypesemdiff{1}, 
\archetypesemdiff{2}, 
\archetypesemdiff{3}, 
\archetypesemdiff{4}, 
\archetypesemdiff{5}, 
and
and \archetypesemdiff{6}.

In recent work, we used archetypometrics to find that canonically female characters tended more toward \archetype{Hero}, \archetype{Adventurer}, \archetype{Diva}, and \archetype{Sophisticate} archetypes, while male characters tended toward \archetype{Fool}, \archetype{Traditionalist}, \archetype{Brute} and \archetype{Outcast} types. On the whole, we found that overarching patterns by gender nevertheless sustained traditional stereotypes: the seemingly positive heroic bias of females was undercut by the correlation between heroism and masculinity in female characters~\cite{beauregard2026archetypesgenderfictiondatadriven}.

We (1) briefly review the relevant history of vampiric myths, focusing on their female counterparts; (2) describe our methods, providing background on the archetypometric dataset, approach, and additional analytical methods used here; (3) discuss our findings and limitations, grounding them in existing theory and outlining directions for future work. 

\section{Literature Review}
\label{sec:buffy-versus-bella.literature-review}

Vampire narratives are strikingly diverse globally, and the origins of our contemporary, blood-mad creatures of the night have passed through a series of cultures and etymologies to arrive at the seductive, modern-day hematophage. These stories cross borders, and yet, recently follow a pattern of depiction. For example, hungry, reanimated corpses appear in Chinese mythologies as early as the Qing Dynasty, but the so-called jiangshi have fused with characteristics of Western vampires~\cite{ancuta2023}, a sort of magnetism that, as we will see, speaks to how vampires mirror society's concerns.

First, we examine the manner in which vampires reflect bodily fears. In the Western hemisphere, the vampire became an obsession of the Romantic movement in literature. Nineteenth century European vampires fed not on blood, but on psychic energy~\cite{bartholomew2024}, merging with the Romantics' moral panic and fascination with tuberculosis~\cite{fenn2021}, a wasting disease that slowly unspooled the life from its victims. John Polidori published his short story, ``The Vampyre''~\cite{polidori1819vampyre}, 
in 1819 based on his medical theses~\cite{stiles2010,tiziani2009}, drawing a parallel between vampirism and medical scholarship that reinforced the notion of the `beast of the night' as representing a deadly and terrible infliction of the time.

In Bram Stoker's ``Dracula''~\cite{stoker1897dracula}
the vampire loses their association with disease and becomes a highborn, predatory character.
Groom \cite{groom2018} likens Stoker's Count Dracula as the all-encompassing ``black hole'' of vampire mythos, coining the term ``After Dracula (AD)'' to sift through the substantial body of literature that succeeded Stoker. In crafting Dracula, Stoker, too, relied on moral panic: Parts of the novel were said to be based on the case of Jack the Ripper~\cite{rance2002}. 
The manner in which depictions of violence and disease were morphed into sexuality has been of interest to scholars like Kimberly Frohreich \cite{frohreich2013}, who views the shift to sexuality as an intermingling of threat with seduction. 

More broadly, vampires serve as a mirror for powerful forces in (or around) society~\cite{Auerbach1995} 
that control and victimize the populace. 
Such a fear is entirely reasonable since living in society means compromising between our own inclinations and collective needs, the latter of which can be overwhelmed by those who are powerful. Vampires represent the powerful who have leveraged that to their advantage; that is why vampiric metaphors have applied to people or groups seen as exploiting an opportunity~\cite{Marx1867WorkingDay}. Karl Marx mentions vampires in his seminal 1867 work \textit{Capital} \cite{Marx1867WorkingDay}:  ``Capital is dead labour, that, vampire-like, only lives by sucking living labour, and lives the more, the more labour it sucks. The time during which the labourer works, is the time during which the capitalist consumes the labour-power he has purchased of him.'' Death, vampirism, consumption, wealth, life, sucking, and exploitation intertwine in additional passages: ``It quenches only in a slight degree the vampire thirst for the living blood of labour'' and ``the vampire will not lose its hold on him `so long as there is a muscle, a nerve, a drop of blood to be exploited'''~\cite{Marx1867WorkingDay}. 

In the late 20th century, vampires experienced a surge in popularity in fiction, film, and television, particularly among teenage audiences.
Basu (2018) attributes this to the fusion of traditional horror with fantasy genres~\cite{basu2018}. 
Beginning in the 1990s, 
film adaptations of vampire novels, 
including \textit{Dracula} (1992)
and 
Anne Rice's \textit{Interview with the Vampire} (1994), 
made their way into theaters.
Comedy-horror versions of vampire stories 
also emerged around this time
with 
\textit{The Lost Boys} (1987)
and
\textit{Buffy the Vampire Slayer} (1992),
the latter leading to 
the same-named television series (1997--2003).
Buffy the Vampire Slayer would become especially influential 
in popular culture~\cite{wilcox2005buffy,stevenson2004,wilcox_lavery2002}.
For one example, the 
many creative, playful, and subversive
elements and storylines of BtVS 
were the direct inspiration for
\href{https://tvtropes.org}{tvtropes.org}~\cite{tvtropes}.

And while still dangerous, 
vampires 
began to take on deeply 
desirable and aspirational appeal for fans.
From \textit{The Lost Boys}: 
``Sleep all day, party all night, never grow old, never die. It's fun to be a vampire''~\cite{schumacher1987lostboys}.
In BtVS, 
Angel 
and 
Spike, idolized as romantic rivals, represented the spectrum of male desirability.  
For \textit{Twilight}, fans attended midnight premiers of movies wearing t-shirts with Team Edward and Team Jacob. 
In \textit{Vampire Diaries}, brothers Damon and Stefan Salvatore fight for the affections of Elena Gilbert, much to the jealousy of her fellow students. 

While Joseph Le Fanu's 1872 novella \textit{Carmilla} featured lesbian, empowered female vampires~\cite{Ridenhour2013}, the representation of women vampire stories remained stagnant (and decidedly male-centered) until the rise of feminism in the mid 20th century~\cite{pender2004}. From this point, female character representation in vampire stories began to expand in both scope and depth~\cite{pender2004}. Mercer \cite{mercer2011} posits that in the popular \storylinksimple{Twilight}{Twilight series}, the lead vampire, Edward, is reconfigured from a violent creature to a ``spiritually attractive figure,'' a so-called vegetarian, in response to changing mores. 
Going further, Luksza \cite{luksza2014} explains the rise in vampire media from a feminist lens, concluding that series such as 
\storylinksimple{Buffy-the-Vampire-Slayer}{Buffy the Vampire Slayer} 
and
\storylinksimple{Twilight}{Twilight} 
represent a shift in Western sexual politics guided by changes in power relations between genders and the sexual objectification of, against years of vampire history, men instead of women.

In the online era, fandoms gained narrative power over vampire stories---and media in general. 
Wang ~\cite{wang2020} asserts that technology has pushed the fan to the ``central stage'' of audience participation in popular culture, giving them direct power of the consumption of materials. Others liken the cult of fandom to a religion~\cite{blom2013}. When it comes to vampires, the call by fandoms to reanalyze a classic text for the modern era has led to women seeking empowerment through the vampire narrative (see Williamson~\cite{williamson2001} for an in-depth investigation). 
Nonetheless, fans frequently imitate fashion~\cite{stenger2006}, molding the world to their own bodies, something especially poignant when we consider the body horror that has historically defined the vampire. 
Arguably, the vampire fandom surrounding \textit{Twilight} and other media is about reclaiming centuries of previous impressions with the present social mores. 
At poles of these mores are the main characters in \textit{Buffy the Vampire Slayer} and \textit{Twilight}.

\section{Description of data sets}
\label{sec:buffy-versus-bella.data}

Our discovery of the archetypometrics framework
drew on a dataset collected from the Open Source Psychometrics Project's~\cite{openpsychometrics} 
``Statistical `Which Character' Personality Quiz.'' 
Participants rated 2,000 fictional characters drawn from 341 stories along 464 semantic differential traits (e.g. \traitlinksimple{orderly}{chaotic}) using a 101-point slider. 
Using singular value decomposition (SVD) of over 72 million ratings,
we uncovered six essential archetype dimensions, as previously mentioned---three primary:
\begin{itemize}
  \item[] \archetypesemdiff{1},
  \item[] \archetypesemdiff{2},
  \item[] \archetypesemdiff{3},
\end{itemize}
\newpage
and three secondary: 
\begin{itemize}
  \item[] \archetypesemdiff{4},
  \item[] \archetypesemdiff{5},
  \item[] \archetypesemdiff{6}.
\end{itemize}
Together, these six dimensions account for 74.6\% of the variance among characters,
and 99.2\% of characters have their largest component along 
one of these dimensions.

\section{Methods}
\label{sec:buffy-versus-bella.methods}

We analyze two female protagonists, \characterlinksimple{Buffy-the-Vampire-Slayer-Buffy-Summers}{Buffy Summers} (\storylinksimple{Buffy-the-Vampire-Slayer}{Buffy the Vampire Slayer}) and \characterlinksimple{Twilight-Bella-Swan}{Bella Swan} (\storylinksimple{Twilight}{Twilight}), within the archetypometric framework, using three levels of comparison: (i) positioning each character within the full 2,000-character dataset, (ii) situating each within the cast of her own story, and (iii) directly contrasting the two characters.

To establish each character's position in archetype space, we use 
``character cards'' (see §4.3 in~\cite{dodds_2025_17128112_pragmateia_archetypes}), which summarize a character's decomposition across the six essential archetype dimensions using normalized component sizes and percentage variance explained, along with overall size rank and archetype ratio within the full dataset. 
The cards also include the most defining traits and nearest neighboring characters in trait space, enabling quantitative comparison of \characterlinksimple{Buffy-the-Vampire-Slayer-Buffy-Summers}{Buffy} and \characterlinksimple{Twilight-Bella-Swan}{Bella} in terms of their archetypal profiles and relative positions within the broader character landscape.

We further analyze character distributions within each story. We use the ousiogram card framework (see §5 in~\cite{dodds_2025_17128112_pragmateia_archetypes}), which visualizes how a narrative's characters are positioned in archetype space. Characters are projected onto pairwise planes of the six essential archetype dimensions (15 total), with labels indicating component magnitude. This approach reveals clustering patterns, dominant archetypes, and the balance between opposing character types within a story, allowing us to assess how each protagonist relates to her surrounding cast.

Finally, we compare \characterlinksimple{Buffy-the-Vampire-Slayer-Buffy-Summers}{Buffy} and \characterlinksimple{Twilight-Bella-Swan}{Bella} in trait space, employing the allotaxonograph framework (see §6 in~\cite{dodds_2025_17128112_pragmateia_archetypes}), which provides an interpretable decomposition of similarity between any two characters in high-dimensional trait space. Each character is encoded as a vector of normalized trait scores, and similarity is computed via the inner product of their trait vectors, a measure invariant to basis choice and robust to the orientation of individual semantic differentials. Allotaxonographs combine a histogram of trait contributions with a ranked list of traits driving similarity or dissimilarity by absolute magnitude, allowing us to identify the specific trait-level mechanisms underlying character discordance beyond a single summary score.

\section{Results}
\label{sec:buffy-versus-bella.results}

Here, we analyze 
\characterlinksimple{Buffy-the-Vampire-Slayer-Buffy-Summers}{Buffy} 
and 
\characterlinksimple{Twilight-Bella-Swan}{Bella} through the three lenses previously mentioned: (i) each character within the full 2,000-character dataset, (ii) each character within the cast of her own story, and (iii) direct contrast between the two characters.

\subsection{The archetypes of Buffy and Bella within a firmament of 2,000 characters}

Characters are represented by a vector in a 464-dimensional trait space.
Character size is the normalized magnitude of the
character's trait vector.
\characterlinksimple{Buffy-the-Vampire-Slayer-Buffy-Summers}{Buffy} rates $646/2000$ in relative character size as opposed to \characterlinksimple{Twilight-Bella-Swan}{Bella} who rates $1388/2000$. 
\characterlinksimple{Buffy-the-Vampire-Slayer-Buffy-Summers}{Buffy's} archetypal rating is relatively large compared to the rest of the dataset; \characterlinksimple{Twilight-Bella-Swan}{Bella's} character size is relatively small. 

\begin{figure*}[t!]
  \centering	
    \includegraphics[width=\textwidth]{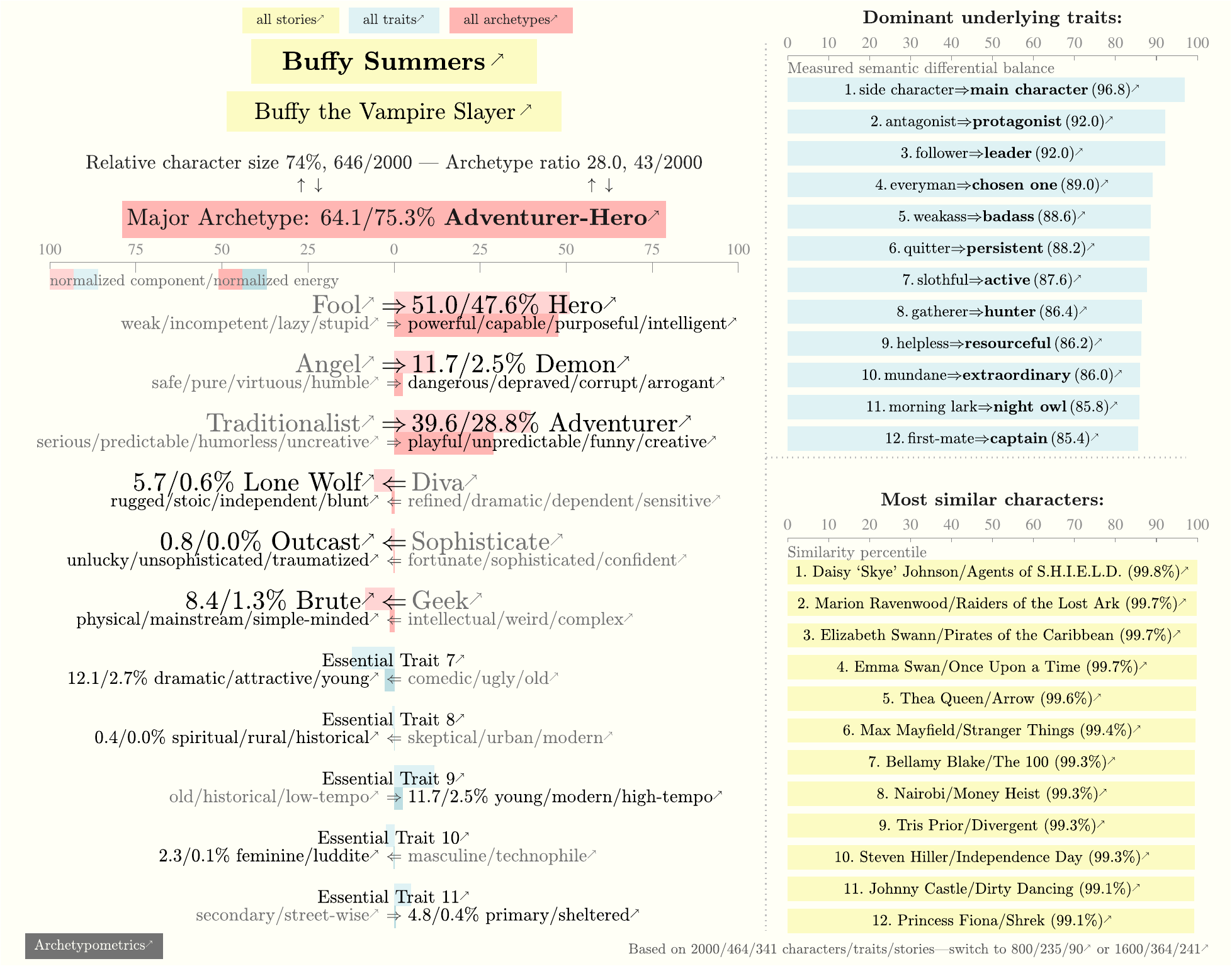}
  \caption{
Character card for \characterlinksimple{Buffy-Summers-Buffy-the-Vampire-Slayer}{Buffy Summers} from \storylinksimple{Buffy-the-Vampire-Slayer}{Buffy-the-Vampire-Slayer}. The left panel depicts the character's relative size in Archetype space, as well as each normalized component of the primary archetypes, secondary archetypes, and essential trait dimensions. \characterlinksimple{Buffy-Summers-Buffy-the-Vampire-Slayer}{Buffy Summers} has a major archetype of \archetypelinksimple{Adventurer-Hero}{Adventurer-Hero}. The right panel depicts her dominant underlying traits, and most similar characters.
  }
  \label{fig:buffy-versus-bella.buffy-card}
\end{figure*}

In Figure~\ref{fig:buffy-versus-bella.buffy-card}, we observe that \characterlinksimple{Buffy-the-Vampire-Slayer-Buffy-Summers}{Buffy} has a Major Archetype of 
\archetypelinksimple{Angel-Hero}{Angel-Hero},
in which her primary archetype is \archetype{Hero} with a strong inclination toward a secondary archetype of \archetype{Adventurer}. The normalized components of her archetype pairing describes $75.8\%$ of her trait vectors. Her most similar characters are female protagonists from adventure and science fiction franchises, like \characterlinksimple{Agents-of-SHIELD-Daisy-Skye-Johnson}{Daisy `Skye' Johnson} from \storylinksimple{Agents-of-SHIELD}{Agents of S.H.I.E.L.D.}, \characterlinksimple{Raiders-of-the-Lost-Ark-Marion-Ravenwood}{Marion Ravenwood} from \storylinksimple{Raiders-of-the-Lost-Ark}{Raiders of the Lost Ark}, and \characterlinksimple{Pirates-of-the-Caribbean-Elizabeth-Swann}{Elizabeth Swann} from \storylinksimple{Pirates-of-the-Caribbean}{Pirates of the Caribbean}. Her dominant underlying traits reflect these similarities: her major traits are 
\traitlinksimplerightalt{side-character}{main-character}{side character}{main character}, 
\traitlinksimpleright{antagonist}{protagonist}, 
and \traitlinksimpleright{follower}{leader} 
(bold text indicates in the direction of main character, protagonist, and leader, a convention we will follow throughout these results). When exploring her trait neighbors---that is, other characters in the dataset with whom she shares similar ratings on her top traits---a slightly different context emerges. In Table~\ref{tab:top_trait_main_neighborhood_list}, we observe that prototypical ``male leader'' characters, like \characterlinksimple{Star-Trek-The-Next-Generation-Jean-Luc-Picard}{Captain Jean-Luc Picard} from \storylinksimple{Star-Trek-The-Next-Generation}{Star Trek: The Next Generation}, dominate the ``main character'' trait rating. \characterlinksimple{Buffy-the-Vampire-Slayer-Buffy-Summers}{Buffy} ranks 4th of all characters in the dataset on ``main character'' trait ratings.

\begin{figure*}[t!]
  \centering	
    \includegraphics[width=\textwidth]{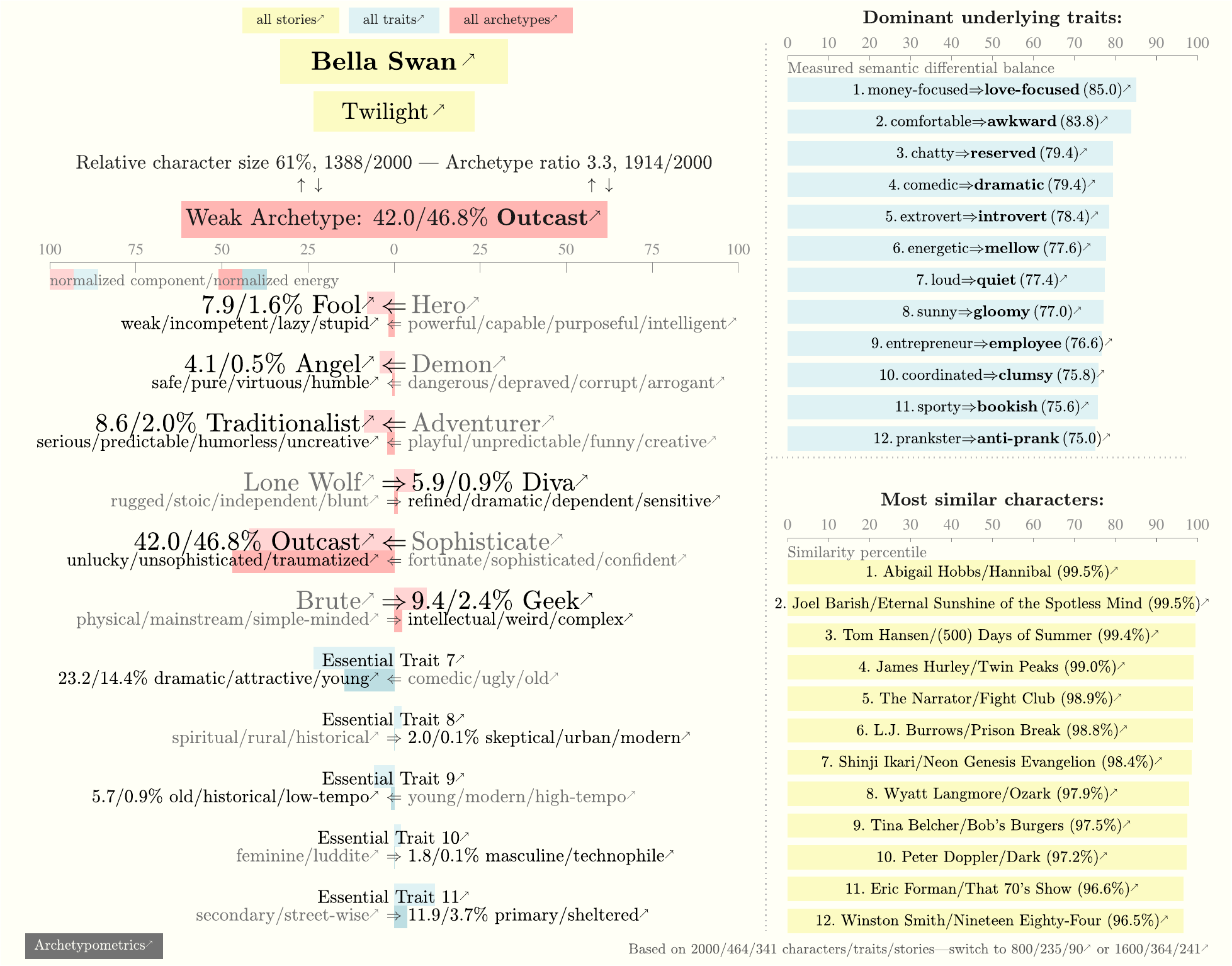}
  \caption{
    Character card for \characterlinksimple{Twilight-Bella-Swan}{Bella Swan} from \storylinksimple{Twilight}{Twilight}. The left panel depicts the character's relative size in Archetype space, as well as each normalized component of the primary archetypes, secondary archetypes, and essential trait dimensions. \characterlinksimple{Twilight-Bella-Swan}{Bella-Swan} has a weak archetype of \archetype{Outcast}. The right panel depicts her dominant underlying traits, and most similar characters.
  }
  \label{fig:buffy-versus-bella.bella-card}
\end{figure*}

In Figure~\ref{fig:buffy-versus-bella.bella-card}, \characterlinksimple{Twilight-Bella-Swan}{Bella} has a Weak Archetype; that is, the magnitudes of her dominant trait vectors relative to other characters in the dataset are relatively small. However, while only $46.8\%$ of her trait vectors point in the direction \archetype{Outcast}, they do not point substantively in the direction of any other primary or secondary archetype pairings. Interestingly, her top three most similar characters, \characterlinksimple{Abigail-Hobbs}{Abigail Hobbs} from \storylinksimple{Hannibal}{Hannibal}, \characterlinksimple{Joel-Barish}{Joel Barish} from \storylinksimple{Eternal-Sunshine-of-the-Spotless-Mind}{Eternal Sunshine of the Spotless Mind}, and \characterlinksimple{Tom-Hansen}{Tom Hansen} from \storylinksimple{500-Days-of-Summer}{(500) Days of Summer}, are also \archetype{Outcast} archetypes, but range in strength from (weak to minor to major, respectively) and also all contain secondary archetypes. For example, Abigail Hobbs is a 
\archetypelinksimple{Geek-Hero}{Geek-Hero}.

\characterlinksimple{Twilight-Bella-Swan}{Bella's} underlying traits are 
\traitlinksimpleright{money-focused}{love-focused}, 
\traitlinksimpleright{charming}{awkward}, 
and 
\traitlinksimpleright{chatty}{reserved}. 
In Table~\ref{tab:top_trait_love_neighborhood_list}, \characterlinksimple{Twilight-Bella-Swan}{Bella's} top trait neighbors for \traitlinksimpleright{money-focused}{love-focused} 
are \characterlinksimple{Anna}{Anna} from \storylinksimple{Frozen}{Frozen}, 
\characterlinksimple{West-Side-Story-Maria}{Maria} 
from \storylinksimple{West-Side-Story}{West Side Story}, 
and \characterlinksimple{Frozen-Olaf}{Olaf} from \storylinksimple{Frozen}{Frozen}: 
all characters with dominant familial and romantic love-based story lines in their respective films. 

\subsection{An archetypometric comparison of Buffy and Bella}

\begin{figure*}[t!]
  \centering	
    \includegraphics[width=\textwidth]{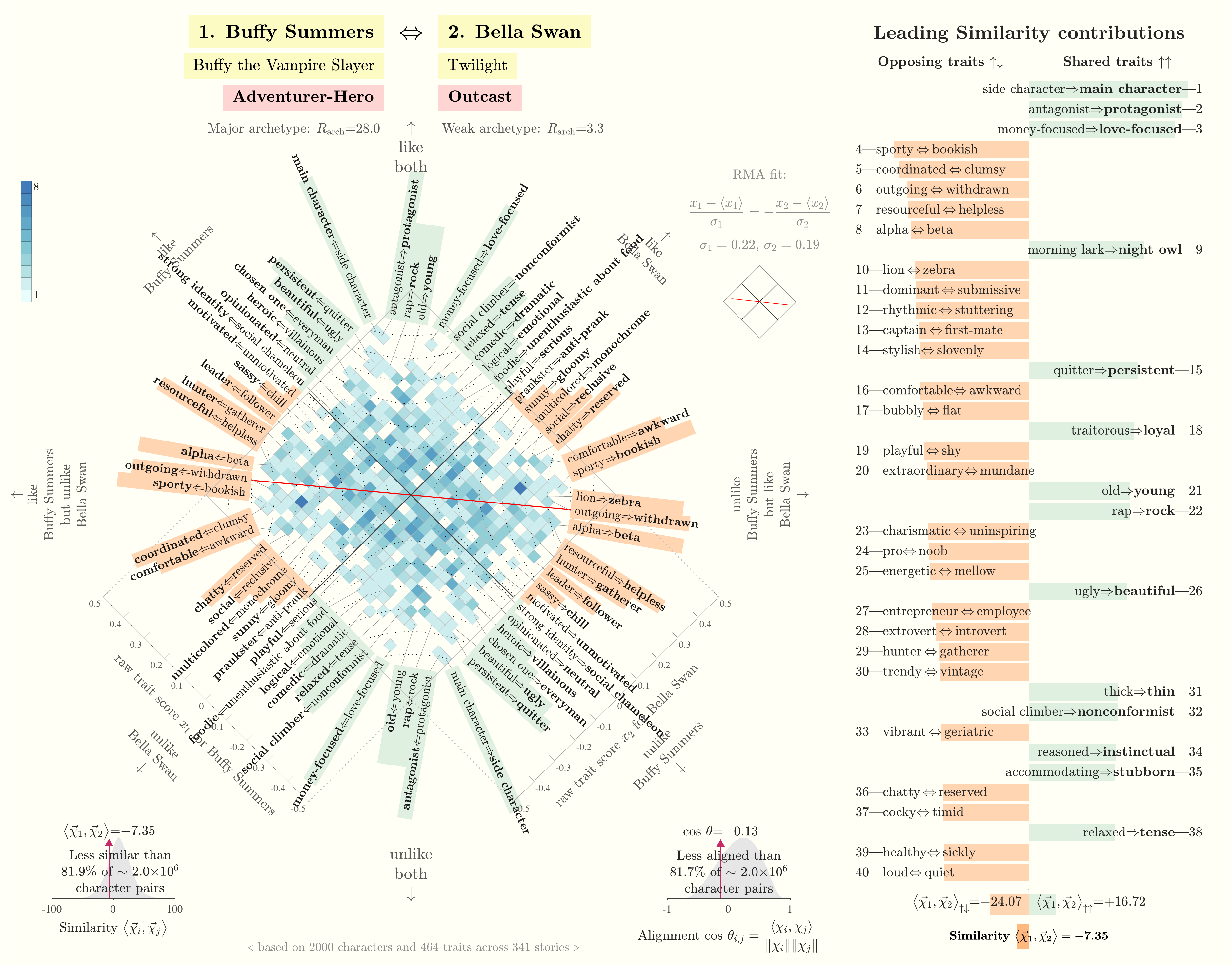}
  \caption{
    Ousiogram card comparing the most similar and most dissimilar traits of \characterlinksimple{Buffy-the-Vampire-Slayer-Buffy-Summers}{Buffy Summers} versus \characterlinksimple{Twilight-Bella-Swan}{Bella Swan}. The right panel depicts opposing traits (that is, traits that contribute most to dissimilarity between the characters) and shared traits, in order of contribution. These characters are reasonably dissimilar, demonstrated by more opposing traits, and a wide Ousiogram appearance. 
  }
  \label{fig:buffy-versus-bella.comparison}
\end{figure*}

Based on pure archetype alone, \characterlinksimple{Buffy-the-Vampire-Slayer-Buffy-Summers}{Buffy} and \characterlinksimple{Twilight-Bella-Swan}{Bella} differ greatly on the strength and variety of their respective archetypes. \characterlinksimple{Buffy-the-Vampire-Slayer-Buffy-Summers}{Buffy} possesses a major archetype pairing, in \archetypelinksimple{Adventurer-Hero}{Adventurer-Hero} and \characterlinksimple{Twilight-Bella-Swan}{Bella} has a weak \archetype{Outcast} archetype. Whereas they differ individually in strength of main traits relative to the total dataset, they share directionality of some traits. In Figure~\ref{fig:buffy-versus-bella.comparison}, they share \traitlinksimpleright{side character}{main character}, \traitlinksimpleright{antagonist}{protagonist}, and \traitlinksimpleright{money-focused}{love-focused} as their top three shared traits (e.g. the traits for which they have the most similar ratings across the full dataset). However, they differ across traits like \traitlinksimpleright{bookish}{sporty}, \traitlinksimpleright{clumsy}{coordinated}, and \traitlinksimpleright{withdrawn}{outgoing}, with \characterlinksimple{Buffy-the-Vampire-Slayer-Buffy-Summers}{Buffy}---who is an aspiring high school cheerleader---displaying prototypical directionality for these traits (in the direction of sporty, coordinated, and outgoing). 

\subsection{Buffy and Bella compared to their own stories}

\begin{figure*}[t!]
  \centering	
    \includegraphics[width=\textwidth]{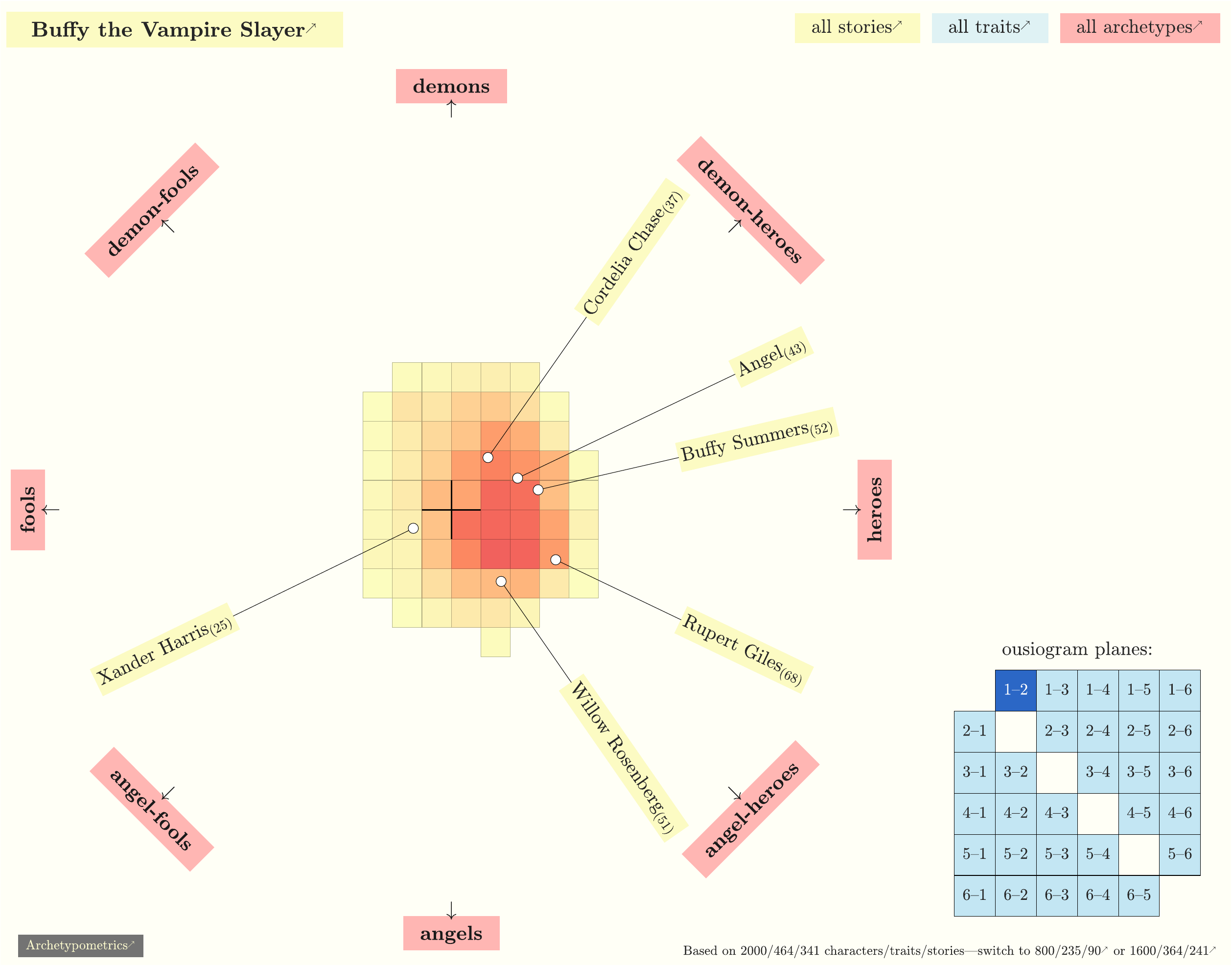}
  \caption{
    All \storylinksimple{Buffy-the-Vampire-Slayer}{Buffy the Vampire Slayer} characters in the archetypometrics dataset across the first and second Archetype dimensions: 
    \protect\archetypesemdiff{1} and \protect\archetypesemdiff{2}. The center of distributions (where characters are weakly present in these dimensions) is depicted by $\mathbf{+}$; 
    the full dataset has a skew towards \archetype{Hero} and \protect\archetype{Angel}, depicted by darker shades of red in the magma color scale. All characters skew toward \archetype{Hero} in \protect\storylinksimple{Buffy-the-Vampire-Slayer}{Buffy the Vampire Slayer}, except \protect\characterlinksimple{Buffy-the-Vampire-Slayer-Xander-Harris}{Xander Harris}, a \protect\archetype{Fool}.
    }
  \label{fig:buffy-characters.comparison}
\end{figure*}

\begin{figure*}[t!]
  \centering	
    \includegraphics[width=\textwidth]{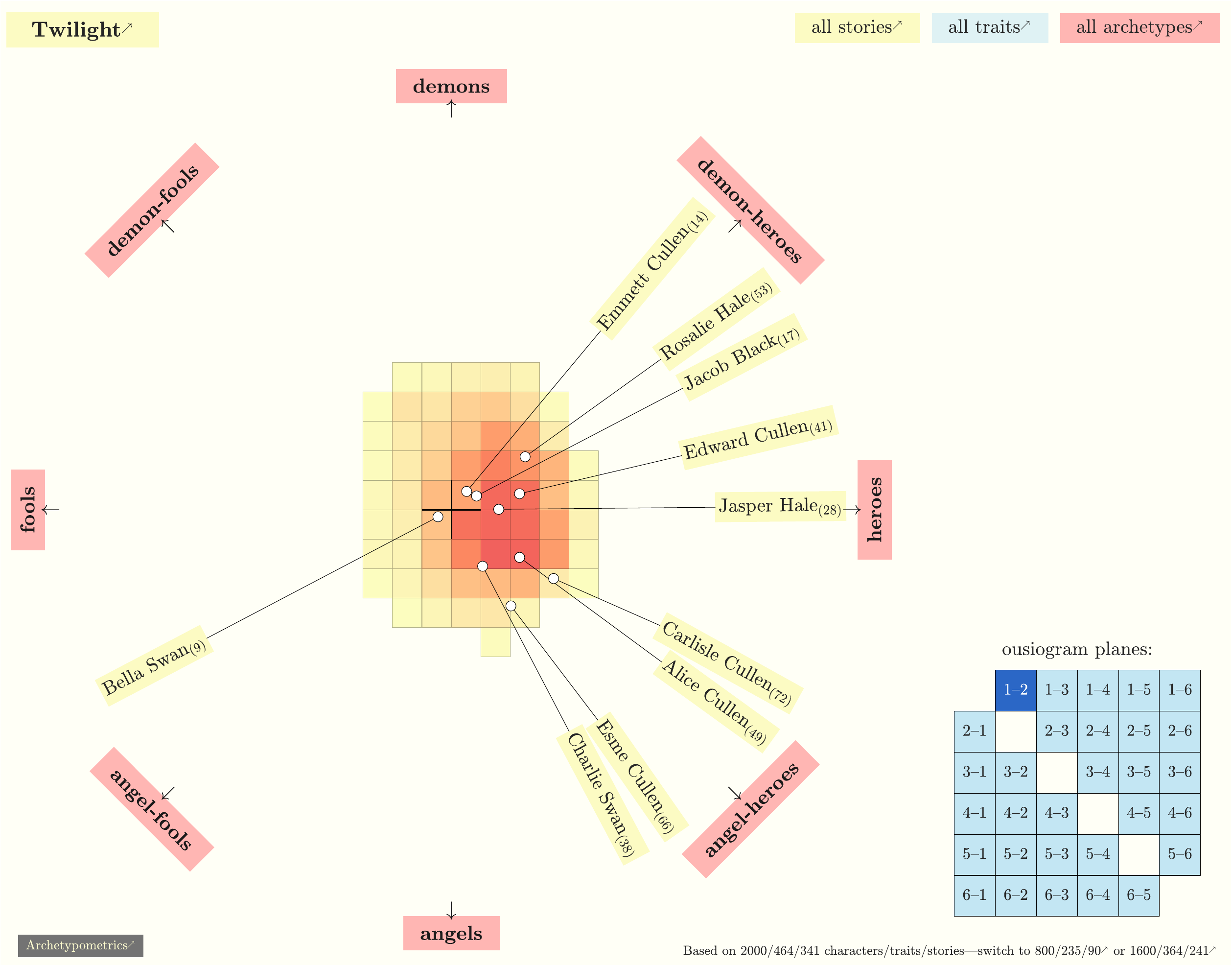}
  \caption{
   All \protect\storylinksimple{Twilight}{Twilight} characters in the archetypometrics dataset across the first and second Archetype dimensions: \protect\archetypesemdiff{1} and \protect\archetypesemdiff{2}. The center of distributions (where characters are weakly present in these dimensions) is depicted by \textbf{+}; the full dataset has a skew towards \archetype{Hero} and \archetype{Angel}, depicted by darker shades of red in the magma color scale. All characters skew \protect\archetype{Hero}, except \protect\characterlinksimple{Twilight-Bella-Swan}{Bella Swan} who skews slightly in the direction \protect\archetype{Fool}. 
   }
  \label{fig:bella-characters.comparison}
\end{figure*}

\characterlinksimple{Buffy-the-Vampire-Slayer-Buffy-Summers}{Buffy} and \characterlinksimple{Twilight-Bella-Swan}{Bella} retain different relationships to their stories. In Figure~\ref{fig:buffy-characters.comparison}, we observe that \characterlinksimple{Buffy-the-Vampire-Slayer-Buffy-Summers}{Buffy} is relatively close to others in her story, and near to her primary love interest,  \characterlinksimple{Buffy-the-Vampire-Slayer-Angel}{Angel}. In Figure~\ref{fig:bella-characters.comparison}, we observe that \characterlinksimple{Twilight-Bella-Swan}{Bella} occupies a more distant position from her fellow characters, most of whom trend towards \archetype{Hero}, with a spread among \archetypesemdiff{2}. \characterlinksimple{Twilight-Bella-Swan}{Bella} is more distant from her primary love interest, \characterlinksimple{Twilight-Edward-Cullen}{Edward Cullen}. Interestingly, we explore the cosine similarity between \characterlinksimple{Buffy-the-Vampire-Slayer-Buffy-Summers}{Buffy} and \characterlinksimple{Buffy-the-Vampire-Slayer-Angel}{Angel}, and \characterlinksimple{Twilight-Bella-Swan}{Bella} and \characterlinksimple{Twilight-Edward-Cullen}{Edward} in Table~\ref{tab:cosine_couples} and note that \characterlinksimple{Buffy-the-Vampire-Slayer-Buffy-Summers}{Buffy} and \characterlinksimple{Twilight-Bella-Swan}{Bella} are dissimilar (with a score of $-0.13$) but \characterlinksimple{Buffy-the-Vampire-Slayer-Angel}{Angel} and \characterlinksimple{Twilight-Edward-Cullen}{Edward} are similar (with a score of $0.79$). 

\begin{figure*}[t!]
  \centering	
    \includegraphics[width=\textwidth]{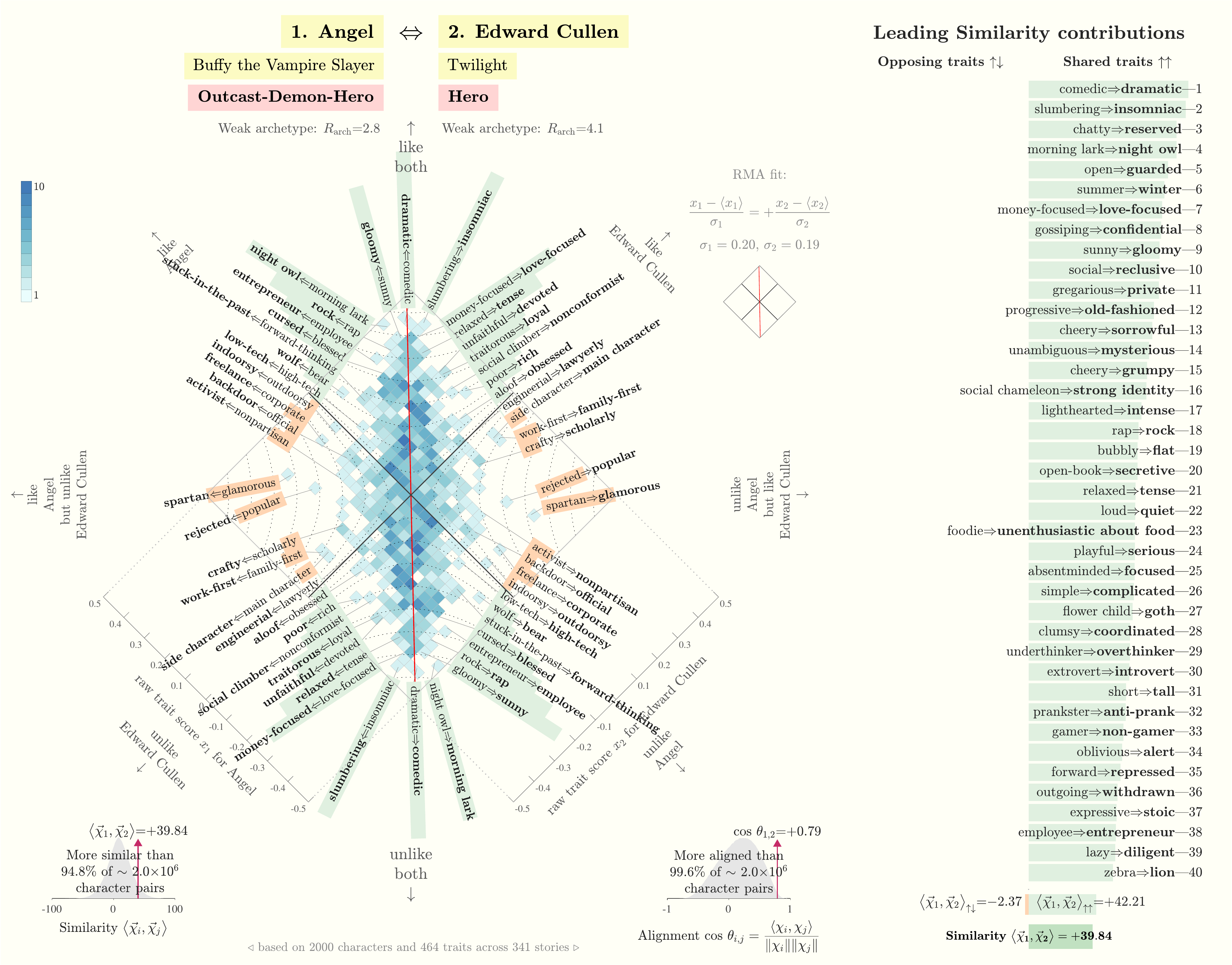}
  \caption{
        Ousiogram card comparing the most similar and most dissimilar traits of \characterlinksimple{Buffy-the-Vampire-Slayer-Angel}{Angel} versus \characterlinksimple{Twilight-Edward-Cullent}{Edward Cullen}. The right panel depicts opposing traits (that is, traits that contribute most to dissimilarity between the characters) and shared traits, in order of contribution. These characters are highly similar to one another, demonstrated by a preponderance of shared traits by contribution, as well as a reasonably narrow Ousiogram. 
  }
  \label{fig:angel-versus-edward.comparison}
\end{figure*}

To further illustrate these similarities, Figure~\ref{fig:angel-versus-edward.comparison} shows that \characterlinksimple{Buffy-the-Vampire-Slayer-Angel}{Angel} and \characterlinksimple{Twilight-Edward-Cullen}{Edward} strongly share $40$ top traits ranging from \traitlinksimpleright{comedic}{dramatic} (1st most similar) to \traitlinksimple{sunny}{gloomy} (9th) and \traitlinksimple{relaxed}{tense} (21st).

Additionally, we note that \characterlinksimple{Buffy-the-Vampire-Slayer-Buffy-Summers}{Buffy's} and \characterlinksimple{Twilight-Bella-Swan}{Bella's} male parental figures markedly diverge in archetype space.  \characterlinksimple{Buffy-the-Vampire-Slayer-Buffy-Summers}{Buffy's} \characterlinksimple{Buffy-the-Vampire-Slayer-Buffy-Summers}{Rupert Giles} has a minor \archetypelinksimple{Angel-Hero}{Angel-Hero} and \characterlinksimple{Twilight-Bella-Swan}{Bella's} \characterlinksimple{Twilight-Charlie-Swan}{Charlie Swan} has a weak three-part archetype of \archetypelinksimple{Outcast-Traditionalist-Angel}{Outcast-Traditionalist-Angel}.

\begin{table}[t!]
    \centering
    \begin{tabular}{p{3.6cm} p{1.85cm} p{1.85cm}}
        \hline
         & \textbf{Bella} & \textbf{Buffy} \\
        \hline
        Distance from lead (all) & 5.93 (0.96) & 5.39 (1.06) \\
        Distance from own & 4.05 (0.37) & 3.61 (0.33) \\
        Distance from other & 6.14 (0.66) & 6.12 (0.95) \\
        Similarity to lead (all) & 0.02 (0.20) & 0.31 (0.26) \\
        Similarity to own & 0.43 (0.12) & 0.69 (0.06) \\
        Similarity to other & -0.11 (0.15) & -0.03 (0.25) \\
        Neighbors' $R_{g}$ & 1.74 & 2.03 \\
        Story dimensions' $R_{g}$ & 0.64 & 0.53 \\
        \hline
    \end{tabular}
    \caption{Distance and similarity metrics for \characterlinksimple{Buffy-the-Vampire-Slayer-Buffy-Summers}{Buffy} and \characterlinksimple{Twilight-Bella-Swan}{Bella} comparisons in archetype space. Values except radius of gyration ($R_{g}$) include `$\mu$~($\sigma$)'. Comparisons include lead characters': distance from all (2000), distance from their neighbors, distance from the other lead's neighbors; similarity to all (2000), similarity to their neighbors, similarity to the other lead's neighbors; and $R_{g}$ for their neighbors, story group's archetype dimensions, and story group's traits. Lower distance has a similar interpretation to higher similarity.}
    \label{tab:distance_sim_summary}
\end{table}

\section{Discussion}

To contextualize these results, we can look to \characterlinksimple{Buffy-the-Vampire-Slayer-Buffy-Summers}{Buffy} and \characterlinksimple{Twilight-Bella-Swan}{Bella's} stories and understand that they represent a refutation and an adoption of a traditionalist, Western conception of femininity, respectively. There are three areas where this distinction is most visible: the characters' orientation to romantic love, their physicality and approach to violence, and their place among the larger cast of their stories.

First, the characters differ significantly with respect to the role of romantic love in their stories. The \traitlinksimpleright{money-focused}{love-focused} trait does not appear in the list of \characterlinksimple{Buffy-the-Vampire-Slayer-Buffy-Summers}{Buffy's} top 12, but is \characterlinksimple{Twilight-Bella-Swan}{Bella's} most dominant underlying trait. This makes sense with respect to the stories each character is from: \storylinksimple{Buffy-the-Vampire-Slayer}{Buffy the Vampire Slayer} is a fantasy story with romantic elements, while \storylinksimple{Twilight}{the Twilight Saga} is a romantic story with fantasy elements. 

While \characterlinksimple{Buffy-the-Vampire-Slayer-Buffy-Summers}{Buffy} is involved in romantic story lines throughout \storylinksimple{Buffy-the-Vampire-Slayer}{Buffy the Vampire Slayer}, it is not the main purpose of her character to fall in love and be in a relationship---in fact, romance is often an obstacle to \characterlinksimple{Buffy-the-Vampire-Slayer-Buffy-Summers}{Buffy's} greater goals and can be a source of tension for her character~\cite{pender2004}. In this sense, \characterlinksimple{Buffy-the-Vampire-Slayer-Buffy-Summers}{Buffy} is a less traditionally feminine character, as these conceptions of femininity posit that women be primarily preoccupied and motivated by affection, compassion, and tenderness~\cite{bem1974}. In contrast, \characterlinksimple{Twilight-Bella-Swan}{Bella's} character arc is entirely subsumed by her romance with \characterlinksimple{Twilight-Edward-Cullen}{Edward Cullen}, and most of her feelings, motivations, and actions in the text revolve around their relationship~\cite{stasiewicz-bienkowska2023,mercer2011}. Success for \characterlinksimple{Twilight-Bella-Swan}{Bella} means being able to live happily-ever-after with \characterlinksimple{Twilight-Edward-Cullen}{Edward Cullen}, which makes her orientation towards romantic live fundamentally different from \characterlinksimple{Buffy-the-Vampire-Slayer-Buffy-Summers}{Buffy's}. 

This pattern grows more apparent upon inspection of the particular nature of \characterlinksimple{Buffy-the-Vampire-Slayer-Buffy-Summers}{Buffy's} and \characterlinksimple{Twilight-Bella-Swan}{Bella's} relationships with their love interests.  \characterlinksimple{Buffy-the-Vampire-Slayer-Angel}{Angel} and \characterlinksimple{Twilight-Edward-Cullen}{Edward} are quite similar in Archetype space (with a cosine similarity of $0.79$), whereas \characterlinksimple{Buffy-the-Vampire-Slayer-Buffy-Summers}{Buffy} and \characterlinksimple{Twilight-Bella-Swan}{Bella} are dissimilar (cosine similarity of $-0.13$); this suggests that the story lines approach the relationships differently. Seemingly, the female protagonists are written dissimilarly in comparison to similar Vampiric male leads. 
    
Second, \characterlinksimple{Buffy-the-Vampire-Slayer-Buffy-Summers}{Buffy} and \characterlinksimple{Twilight-Bella-Swan}{Bella} are differentiated by their physicality and appearance. \characterlinksimple{Buffy-the-Vampire-Slayer-Buffy-Summers}{Buffy} is characterized as sporty and coordinated, \characterlinksimple{Twilight-Bella-Swan}{Bella} as bookish and clumsy. These traits, pertaining to perceptions of the characters' physical capacity, can be seen as downstream of their relationship to violence: specifically, battle- or self-defense related fighting. \characterlinksimple{Buffy-the-Vampire-Slayer-Buffy-Summers}{Buffy} is a vampire hunter, so fighting is what she does. \characterlinksimple{Twilight-Bella-Swan}{Bella} does not fight, as a general rule. The instances in the text where she is prepared to fight are motivated by the need to protect her family. This form of violence is the only one that is at all socially acceptable for women under the traditionalist conception of femininity~\cite{huston1985}. A mother can protect her child (this is, in fact, the motivation for \characterlinksimple{Twilight-Bella-Swan}{Bella}'s preparation for war in the final book of the \storylinksimple{Twilight}{Twilight} series), but she cannot be aggressive or warrior-like under traditional norms outside of her immediate role as a protective mother~\cite{hubbard-mattix2019bella}.

Furthermore, \characterlinksimple{Buffy-the-Vampire-Slayer-Buffy-Summers}{Buffy's} outward physicality and fighting capability do not fit neatly within the confines of traditional feminine gender image~\cite{mahalik2005}, whereas \characterlinksimple{Twilight-Bella-Swan}{Bella's} awkwardness and lack of athleticism do. While \characterlinksimple{Buffy-the-Vampire-Slayer-Buffy-Summers}{Buffy} \textit{is} a beautiful cheerleader---which at first seems purely ``traditional''---some scholarship posits that \characterlinksimple{Buffy-the-Vampire-Slayer-Buffy-Summers}{Buffy's} iconic blonde hair is a subversive story-telling mechanism that retains ``her desirab[ility] [and] marketab[ility]'' and helps her ``remain the kind of heroine the era could tolerate''~\cite{lepage2025}. 

In the books and films, \characterlinksimple{Twilight-Bella-Swan}{Bella} supposedly deviates from the norm for her gender by being uninterested in what some would consider trivial feminine pursuits: makeup, fashion, etc. (\characterlinksimple{Twilight-Bella-Swan}{Bella} notably wore Converse to her high school prom~\cite{twilight2008imdb}). While this may seem to be a refutation of traditional gender roles for women, arguably \characterlinksimple{Twilight-Bella-Swan}{Bella} is adhering to a feminine quality which is so often denigrated and disliked in women by men: a superficial infatuation with one's appearance~\cite{Forbes2007}. She does not seek to draw attention to herself, but is instead content to retreat to the background and let more important people take center stage: she is the epitome of modesty in line with traditional norms~\cite{mahalik2005} of femininity.

Bella is in fact shy, quiet, and petite (See Dworkin~\cite{dworkin2001} for an extensive review of traditional conceptions of ``appropriate'', ``small'' female body image). Even her character size in the Archetypes dataset is smaller than \characterlinksimple{Buffy-the-Vampire-Slayer-Buffy-Summers}{Buffy's}. This is critical to \characterlinksimple{Twilight-Bella-Swan}{Bella's} characterization of an innocent in the \storylinksimple{Twilight}{Twilight} series: in both the book series and the movies, \characterlinksimple{Twilight-Edward-Cullen}{Edward} uses the phrase ``And so the lion fell in love with the lamb,'' which refers not only to \characterlinksimple{Twilight-Bella-Swan}{Bella} being literally prey for vampires as a human, but to her personality as a character (helpless, shy, submissive, small). Famously, even the framing of the first film in \storylinksimple{Twilight}{Twilight} positions \characterlinksimple{Twilight-Bella-Swan}{Bella} as weak; it starts with a ``point-of-view camera... sharing the point of view of the predator... hunting down the vulnerable deer''~\cite{larsson2011bookchapter}. 


Lastly, \characterlinksimple{Buffy-the-Vampire-Slayer-Buffy-Summers}{Buffy} and \characterlinksimple{Twilight-Bella-Swan}{Bella} differ in their role among cast members. In terms of interactions with female characters, \characterlinksimple{Buffy-the-Vampire-Slayer-Buffy-Summers}{Buffy} maintains strong female friendships that anchor and expand her powers~\cite{ross2004}. Her friendship with \characterlinksimple{Buffy-the-Vampire-Slayer-Willow-Rosenberg}{Willow Rosenberg} helps her fight evil, remain grounded, and provides the basis for many of her heroic actions. She, unlike \characterlinksimple{Twilight-Bella-Swan}{Bella}, is part of a team, and fills a complementary position in Archetype space, shown in Figure~\ref{fig:buffy-characters.comparison}. 
As discussed previously, \characterlinksimple{Twilight-Bella-Swan}{Bella}, is portrayed as being unlike other girls in her own story, which reflects in her Archetypal position, shown in Figure~\ref{fig:bella-characters.comparison}. In Archetype space, she differs from all other characters with ratings for \storylinksimple{Twilight}{Twilight} as an \archetype{Outcast}.  
Furthermore, while \characterlinksimple{Twilight-Bella-Swan}{Bella} maintains relationships with female counterparts, they are almost entirely though her association with \characterlinksimple{Twilight-Edward-Cullen}{Edward}; all female characters in our dataset for the \storylinksimple{Twilight}{Twilight series} are notably relatives of her lover.  

Further evidence of refutations and reinforcements of traditional gender roles are apparent in the protagonists' relationships with their respective male parental figures. \characterlinksimple{Buffy-the-Vampire-Slayer-Buffy-Summers}{Rupert Giles'} is \characterlinksimple{Buffy-the-Vampire-Slayer-Buffy-Summers}{Buffy's} ``Watcher'', both trainer and confidante in her quest to slay vampires and protect the people of Sunnydale~\cite{buffy1997imdb}. While at times \characterlinksimple{Buffy-the-Vampire-Slayer-Rupert-Giles}{Giles} can be aloof and not keep the Slayer informed of pertinent information (see S2E8 ``The Dark Age'' for bad outcomes from this approach), he trains, supports, and uplifts \characterlinksimple{Buffy-the-Vampire-Slayer-Buffy-Summers}{Buffy} in all her martial, slaying duties~\cite{levy2003buffy}. \characterlinksimple{Twilight-Charlie-Swan}{Charlie Swan}, however, plays a more traditional and paternalistic role as a loving, but largely stereotypical father, who, in a memorable scene, bumblingly tries to give \characterlinksimple{Twilight-Bella-Swan}{Bella} ``the talk'' before she heads out on a date with \characterlinksimple{Twilight-Edward-Cullen}{Edward}. In both cases, the archetypes of these characters support this view: \characterlinksimple{Buffy-the-Vampire-Slayer-Buffy-Summers}{Giles'} is a fairly straightforward \archetypelinksimple{Angel-Hero}{Angel-Hero}, whereas \characterlinksimple{Twilight-Charlie-Swan}{Charlie Swan} possesses a \archetype{Traditionalist} component in his archetype. 

\subsection{Limitations}
Fandom, or ``the community that fans self-constitute around [a] text or object''~\cite{hellekson2015}, around \storylinksimple{Buffy-the-Vampire-Slayer}{Buffy the Vampire Slayer} and \storylinksimple{Twilight}{Twilight} contextualizes our analysis of these two main characters, which may inadvertently amplify and subsume specific nuance. Firstly, our original data source from OpenPsychometrics~\cite{openpsychometrics} is completely crowd-sourced and unpaid. The site stipulates that respondents need to be familiar with the characters they rate, and each survey requires time and effort to respond; it is thus likely that many of the character ratings arise from fans of the shows and films (who may or may not self-identify as part of the ``fandom''). Considering that fandom often begets community norms and self-policing on the internet~\cite{hellekson2015}, our understanding of \characterlinksimple{Twilight-Bella-Swan}{Bella} and \characterlinksimple{Buffy-the-Vampire-Slayer-Buffy-Summers}{Buffy} necessarily nest within both individual as well as group-level psyche. Thus, while critical literature on vampire stories and mythology provide insight into the cultural context of our analysis, the impact of fandom may further remix tropes, perceptions, and responses that underpin our dataset and findings. 

Relatedly, while we attempt to provide a brief, but broad overview of vampire studies, our supporting literature and grounding citations are subject to the social and cultural influence. Euro-centrism and colonialism has shaped the existing scholarship on vampires~\cite{Dalton2024GeopoliticalVampirism}. Our analysis centers these two American stories in a Western cultural context accordingly, but we strongly note that vampires are not unilaterally Western, and our analysis here does not encapsulate all instantiations of vampiric myth and folklore. 

\section{Concluding remarks}
\label{sec:buffy-versus-bella.concludingremarks}
Vampires and their study span the breadth of social and psychological phenomena, both historically and today; \characterlinksimple{Buffy-the-Vampire-Slayer-Buffy-Summers}{Buffy} and \characterlinksimple{Twilight-Bella-Swan}{Bella} are two exemplars by which vampire myth incorporates modern conceptions, and it is clear that outdated and antiquated views of femininity persist through the latter's characterization. Though \storylinksimple{Buffy-the-Vampire-Slayer}{Buffy the Vampire Slayer} predates the \storylinksimple{Twilight}{Twilight} book and film series by a decade, \characterlinksimple{Buffy-the-Vampire-Slayer-Buffy-Summers}{Buffy} is a far stronger female main character. Through archetypometrics, we observe that \characterlinksimple{Buffy-the-Vampire-Slayer-Buffy-Summers}{Buffy} pushes the boundaries of female empowerment and equals \characterlinksimple{Buffy-the-Vampire-Slayer-Angel}{Angel's} \archetype{Hero} tendencies, whereas \characterlinksimple{Twilight-Bella-Swan}{Bella} more traditionally adheres to gender norms centered on her romantic relationship with \characterlinksimple{Twilight-Edward-Cullen}{Edward} even with her \archetype{Outcast} persona. 
More broadly, we showcase how archetypometrics illuminates the complexities of vampire narrative today, showing that modern female protagonists continue to operate in roles that shift and contort alongside society's changing values.

\section*{Acknowledgments}

The authors are grateful for support from the
National Science Foundation (Award \#2242829),
the Massachusetts Mutual Life Insurance Company,
and an anonymous philanthropic donor.

\clearpage

\bibliography{buffy-versus-bella}


\onecolumn

\appendix
\section{Appendices}

\setcounter{page}{1}
\renewcommand{\thepage}{A\arabic{page}}
\renewcommand{\thefigure}{A\arabic{figure}}
\renewcommand{\thetable}{A\arabic{table}}
\setcounter{figure}{0}
\setcounter{table}{0}

\renewcommand{\thesection}{A\arabic{section}}
\setcounter{section}{0}

\section{Neighborhood lists}

The following lists are Bella and Buffy's nearest neighborhoods in archetype space (see Tables~\ref{tab:bella_neighborhood_list} and~\ref{tab:buffy_neighborhood_list}). Their lists are seeded from the top $m$ closest characters to them by euclidean distance before we retrieve those $m$ character's $k$ nearest neighbors as well, building the neighborhood somewhat recursively.

An alternate method to seed initial characters for a neighborhood was to use each lead character's top trait. For Bella, this trait is \traitlinkonelabel{love-focused}{money-focused}{love-focused}. For Buffy, this trait is \traitlinkonelabel{main-character}{side-character}{main-character}. The following tables list the characters in their respective top trait neighborhoods built by capturing the window around their locations in their sorted trait scores, taking $m$ nearest (centered on the lead character) and finding each $m$'s $k$ most similar neighbors in euclidean space (Tables~\ref{tab:top_trait_love_neighborhood_list} and~\ref{tab:top_trait_main_neighborhood_list}). Bella's top trait list overlapped with her neighborhood by $6$ characters while Buffy's top trait list overlapped with her own by $37$ characters.

\clearpage
\begin{table*}[t]
    \centering
    \begin{tabular}{p{4cm} p{4cm} p{4cm}}
        \hline
        1-36 & 37-72 & 73-106 \\
        \hline
        Abigail Hobbs & Caleb Smith & Harry Crane \\
        Joel Barish & Harvey Kinkle & Danny Vinyard \\
        Tom Hansen & Otis Milburn & Piper Chapman \\
        James Hurley & Michael Dawson & Will Graham \\
        The Narrator & Claire Littleton & Heywood \\
        L.J. Burrows & Dan Espinoza & Ray Ploshansky \\
        Shinji Ikari & Connell Waldron & Donna Hayward \\
        Wyatt Langmore & Nick Dunne & Cyclops \\
        Tina Belcher & Julie Taylor & Violet Parr \\
        Peter Doppler & Iceman & Nick Carraway \\
        Eric Forman & Shauna & Nick Blaine \\
        Winston Smith & Kate Pearson & Milhouse Van Houten \\
        Rachel Garrison & Louis Tully & Landry Clarke \\
        Jonathan Byers & Charlie Kelmeckis & Jasper Hale \\
        Vanya Hargreeves & Dean Forester & Laurel Castillo \\
        Stuart Bloom & Stan Marsh & Princess Darya Oblonskaya \\
        Dan Humphrey & Drew Baylor & Michelle \\
        Rogue & Toby Cavanaugh & Alan Harper \\
        Lori Grimes & Felix Lutz & Evan \\
        Chip Dove & Nino Quincampoix & Ray Stantz \\
        Theon Greyjoy & Billy Bibbit & Marie Schrader \\
        Lambert & Jim Burns & Edna Krabappel \\
        Morty Smith & Pete Hornberger & Stu \\
        Elena Gilbert & Guy Montag & Robby Keene \\
        Mike Wheeler & Kermit & Ezra Fitz \\
        Roland Pryzbylewski & Will Byers & Steve Brady \\
        Ted Mosby & Andrea & Artie Abrams \\
        Shaun & Margo Dunne & Kenny McCormick \\
        Flynn White & Abbi Abrams & Lane Pryce \\
        Adam Groff & A.J. Soprano & Leonard Hofstadter \\
        Calvin Weir-Fields & Konstantin Levin & Denny Brosh \\
        Sam Witwicky & Mike McLintock & Kinsey Locke \\
        Dale Harding & Agatha & George Costanza \\
        Tina Cohen-Chang & Bonnie Winterbottom & Moe Szyslak \\
        Jonas Kahnwald & Martha Nielsen &  \\
        Alfredo Linguini & Niko Polastri &  \\
        \hline
    \end{tabular}
    \caption{Bella's nearest neighborhood, sorted by euclidean distance to Bella in archetype space. Position 1 is closest (most similar) to Bella. $N=106$.}
    \label{tab:bella_neighborhood_list}
\end{table*}

\begin{table*}[t]
    \centering
    \begin{tabular}{p{5cm} p{5cm} p{5cm}}
        \hline
        1-36 & 37-72 & 73-106 \\
        \hline
        Daisy 'Skye' Johnson & Sirius Black & Jimmy 'B-Rabbit' Smith Jr. \\
        Marion Ravenwood & Connor MacManus & Sam Button \\
        Elizabeth Swann & Derek Morgan & Katniss Everdeen \\
        Emma Swan & Taystee Jefferson & Robin Buckley \\
        Thea Queen & Wichita & Ygritte \\
        Max Mayfield & Kinsey Locke & Luka Kovac \\
        Bellamy Blake & Robin Scherbatsky & Seeley Booth \\
        Nairobi & Cece Parekh & Lynette Scavo \\
        Tris Prior & Nick Stokes & Rusty Ryan \\
        Steven Hiller & Harry Potter & Ada Shelby \\
        Johnny Castle & June Osborne & Oberyn Martell \\
        Princess Fiona & Brian O'Conner & Lily \\
        Kelly Severide & Laurel Lance & Cassian Andor \\
        Jackson 'Jax' Teller & Elizabeth Bennet & Faye Valentine \\
        Sabrina Spellman & Owen Grady & Dolores Abernathy \\
        Clarke Griffin & Princess Leia & Rhett Butler \\
        Kate Austen & Veronica Lodge & John Connor \\
        Mikaela Banes & Juliana Crain & Heloise \\
        Ciri & Ragnar Lothbrok & Anthony DiNozzo \\
        Daenerys Targaryen & Ekko & James Bond \\
        Mike Lowrey & Bjorn Lothbrok & Raquel Murillo \\
        Toni Topaz & Rey & Rose DeWitt Bukater \\
        David Mills & Dominic Toretto & Tobias 'Four' Eaton \\
        Octavia Blake & Fox Mulder & Eleven \\
        Fiona Gallagher & Nymphadora Tonks & Yennefer \\
        Veronica Fisher & Jo March & Maeve Wiley \\
        Raven Reyes & Claire Randall & Jasmine \\
        Anita & Xiomara Villanueva & Miriam Maisel \\
        Letty Ortiz & Elizabeth Keen & Cheryl Blossom \\
        Sailor Mars & Phil & Butch Coolidge \\
        Jyn Erso & Kiara & Connor Walsh \\
        Gisele Yashar & Sharon 'Boomer' Valerii & Rick Grimes \\
        Wyldstyle & Tokio & Daisy Fuller \\
        Poe Dameron & Oliver Queen & FP Jones \\
        Leela & Roy Harper &  \\
        Captain Marvel & William Riker &  \\
        \hline
    \end{tabular}
    \caption{Buffy's nearest neighborhood, sorted by euclidean distance in archetype space to Buffy. Position 1 is closest (most similar) to Buffy. $N=106$.}
    \label{tab:buffy_neighborhood_list}
\end{table*}

\begin{table*}[t]
    \centering
    \begin{tabular}{p{5cm} p{5cm} p{5cm}}
        \hline
        1-45 & 46-90 & 91-133 \\
        \hline
        Anna & Tami Taylor & Archie Andrews \\
        Maria & Jack Pearson & Bo Peep \\
        Olaf & Claire Randall & Winston Bishop \\
        Sam Obisanya & Beth March & Marmee March \\
        Nomi Marks & Glenn Sturgis & Brian O'Conner \\
        Aang & Cora Munro & Tracy Mills \\
        Mamá Coco & Tom Hansen & Jennifer Parker \\
        Hilda Spellman & Pumbaa & Melanie Hamilton \\
        Charles Boyle & Love Quinn & Ducky \\
        Snow White & Mumen Rider & Rita Hanson \\
        Francis Mulcahy & Princess Anna Karenina & Will Gorski \\
        Juliet Capulet & Flounder & Rita Bennett \\
        Rebecca Bunch & Pam Beesly & Ambrose Spellman \\
        Belle French & Marge Simpson & Doc \\
        Buttercup & Sookie St. James & Ally Maine \\
        \textbf{Bella Swan} & Kimmy Schmidt & Allison Hamilton \\
        Clark Kent & Kristoff & Skylar \\
        Ted Lasso & Debora & Charlie Bucket \\
        Forrest Gump & Héctor & Aragorn \\
        Jamie Fraser & \textbf{Esme Cullen} & Georgiana Darcy \\
        Misa Amane & Noah Calhoun & Manny Santos \\
        Rubeus Hagrid & Erin Hannon & Pop Tate \\
        \textbf{Jacob Black} & Jane Bennet & Helsinki \\
        Anna Bates & Marshall Eriksen & Sam Button \\
        Maid Marian & Theodore Laurence & L.J. Burrows \\
        Romeo Montague & Kara Danvers & Abigail Hobbs \\
        Littlefoot & Frances 'Baby' Houseman & Friar Tuck \\
        Rebecca Pearson & Betty Elms & Kelly Kapoor \\
        Chien-Po & Joel Barish & Ariadne \\
        Cosette & Lorna Morello & Denver \\
        Captain America & Mia Toretto & David Mills \\
        Marianne Dashwood & Darryl Whitefeather & Amy Wong \\
        Miguel & Lisa Johnson & Chloe Sullivan \\
        Rosalind Walker & Rose DeWitt Bukater & John B \\
        David Nolan & Penny & Hannah Horvath \\
        Dorothy Gale & Arthur Weasley & Debbie Gallagher \\
        James Hurley & Belle & Elaine Benes \\
        Eliza Hamilton & Luke Skywalker & Count Alexei Vronsky \\
        Waverly Earp & Benjamin Button & Little John \\
        Sophie Sheridan & Samuel 'Longshot' Anders & Mercutio \\
        Will Turner & Carrie Bradshaw & Kevin Pearson \\
        Aria Montgomery & Annie Reed & Bjorn Lothbrok \\
        Rapunzel & Butters & The Narrator \\
        Mary Margaret Blanchard & Jon Snow &  \\
        Izzie Stevens & Malyen Oretsev &  \\
        \hline
    \end{tabular}
    \caption{Bella's nearest neighbors list built from her location in her top trait, `\textbf{love-focused}`. Characters are sorted by trait value around Bella. $N=132$, and Buffy's position in this trait is $330^{th}$.}
    \label{tab:top_trait_love_neighborhood_list}
\end{table*}

\begin{table*}[t]
    \centering
    \begin{tabular}{p{5cm} p{5cm} p{5cm}}
        \hline
        1-46 & 47-92 & 93-136 \\
        \hline
        Edward Elric & Michael Corleone & John Connor \\
        Jean-Luc Picard & Harvey Specter & Lady Macbeth \\
        Don Draper & Emma Swan & Berlin \\
        \textbf{Buffy Summers} & Lady Mary Crawley & JJ \\
        Maximus & Clarke Griffin & Black Widow \\
        Patrick Jane & Alan Shore & Princess Leia \\
        James Bond & Frank Abagnale & Beth Pearson \\
        Thomas Shelby & Bruce Wayne & Alan \\
        Ender Wiggin & Dominic Toretto & Anna \\
        Indiana Jones & Annalise Keating & Aunt Polly \\
        Olivia Benson & Kat Stratford & Mr. Robot \\
        Dr. Gregory House & Jamie Fraser & Flynn Rider \\
        Sutter Keely & Benjamin Sisko & Linda Belcher \\
        Sailor Moon & Paul Atreides & Wyldstyle \\
        Jason Bourne & Jeff Winger & Sayid Jarrah \\
        Cassie Thomas & Danny Ocean & Perry Cox \\
        Michael Scott & Nathaniel 'Hawkeye' Poe & Melinda May \\
        Kathryn Janeway & Yennefer & Peter Jason Quill \\
        Jackson 'Jax' Teller & Ragnar Lothbrok & Joan Watson \\
        Olivia Pope & Maeve Wiley & Rusty Ryan \\
        John Wick & Robert Langdon & Wichita \\
        Daenerys Targaryen & Connor MacManus & Lena Luthor \\
        Ted Lasso & Clarice Starling & Waverly Earp \\
        Ellen Ripley & David Mills & Sara Sidle \\
        Neal Caffrey & Peter Venkman & Gamora \\
        John McClane & Jo March & Jennifer Jareau \\
        Princess Ariel & Daisy 'Skye' Johnson & Minerva McGonagall \\
        Dolores Abernathy & Owen Grady & Denver \\
        Sabrina Spellman & Christian Grey & Aimee Gibbs \\
        Oliver Queen & Peggy Olson & Lisa Johnson \\
        Jake Peralta & Veronica Lodge & Gavin 'Spinner' Mason \\
        Malcolm Reynolds & Mike Lowrey & Dr. Ian Malcolm \\
        Sherlock Holmes (Sherlock)& Sophie Sheridan & Lily Aldrin \\
        Sherlock Holmes (Elementary)& Johnny Castle & Max Mayfield \\
        Lorraine Broughton & Chuck Bass & George Washington \\
        Simba & Phil & Donkey \\
        Rick Sanchez & Bellamy Blake & Poe Dameron \\
        Rick Grimes & Aragorn & Riff \\
        Hawkeye Pierce & Sun Bak & Marion Ravenwood \\
        Aang & Laura Roslin & Thea Queen \\
        Kate Beckett & Tobias 'Four' Eaton & Bertram Gilfoyle \\
        Stella Gibson & Kate Austen & Peggy Carter \\
        Littlefoot & Rhett Butler & Fez \\
        Maleficent & El Profesor & Bob Pinciotti \\
        June Osborne & Elizabeth Swann &  \\
        Kara Danvers & Atticus Finch &  \\
        \hline
    \end{tabular}
    \caption{Buffy's nearest neighbors list built from her location in her top trait, `\textbf{main character}`. Characters are sorted by trait value around Buffy. $N=135$, and Bella's position in this trait is $350^{th}$.}
    \label{tab:top_trait_main_neighborhood_list}
\end{table*}

\clearpage
\section{Cosine similarity scores for main characters}


\begin{table}[h!]
    \centering
    \begin{tabular}{c c c c c}
        \hline
         & \textbf{Bella} & \textbf{Edward} & \textbf{Buffy} & \textbf{Angel} \\
        \hline
        \textbf{Bella} & 1.00 & 0.31 & -0.13 & 0.23 \\
        \textbf{Edward} & 0.31 & 1.00 & 0.33 & 0.79 \\
        \textbf{Buffy} & -0.13 & 0.33 & 1.00 & 0.47 \\
        \textbf{Angel} & 0.23 & 0.79 & 0.47 & 1.00 \\
        \hline
    \end{tabular}
    \caption{
    Cosine similarity between story pairs Bella and Edward and Buffy and Angel. Higher score means higher similarity (cosine increases as the angle between character vectors decreases).
    }
    \label{tab:cosine_couples}
\end{table}

\end{document}